\documentclass[pop,graphicx,amsmath,amssymb,preprint]{revtex4-2}
\usepackage{graphicx}
\usepackage{comment}
\usepackage{subfig}
\usepackage{esint}
\usepackage{amsthm}
\usepackage{mathtools}

\newtheorem{thm}{Theorem}[section]

\newtheorem{prop}[thm]{Proposition}

\theoremstyle{definition}
\newtheorem{defn}{Definition}[section]

\theoremstyle{definition}
\newtheorem*{defn*}{\protect\definitionname}
\providecommand{\definitionname}{Definition}

\draft 

\begin{document}


\title{Nearly-integrable flows and chaotic tangles in the Dimits shift regime of plasma edge turbulence\footnote{This article may be downloaded for personal use only. Any other use requires prior permission of the author and AIP Publishing. This article appeared in Phys. Plasmas \textbf{30}, 092307 (2023) and may be found at \url{https://doi.org/10.1063/5.0158013}}} 



\author{Norman M. Cao}
\email[]{norman.cao@austin.utexas.edu}
\homepage[]{https://maplenormandy.github.io/}
\affiliation{Courant Institute of Mathematical Sciences, New York University, New York 10012, USA}

\author{Di Qi}
\email[]{qidi@purdue.edu}
\affiliation{Department of Mathematics, Purdue University, West Lafayette, Indiana 47907, USA}


\date{\today}

\begin{abstract}
Transitionally turbulent flows frequently exhibit spatiotemporal intermittency, reflecting a complex interplay between driving forces, dissipation, and transport present in these systems.
When this intermittency manifests as observable structures and patterns in the flow, the characterization of turbulence in these systems becomes challenging due to the nontrivial correlations introduced into the statistics of the turbulence by these structures.
In this work, we use tools from dynamical systems theory to study intermittency in the Dimits shift regime of the flux-balanced Hasegawa-Wakatani (BHW) equations, which models a transitional regime of resistive drift-wave turbulence relevant to magnetically confined fusion plasmas.
First, we show in direct numerical simulations that turbulence in this regime is dominated by strong zonal flows and coherent drift-wave vortex structures which maintain a strong linear character despite their large amplitude.
Using the framework of generalized Liouville integrability, we develop a theory of integrable Lagrangian flows in generic fluid and plasma systems and discuss how the observed zonal flows plus drift waves in the BHW system exhibit a form of ``near-integrability'' originating from a fluid element relabeling symmetry.
We further demonstrate that the BHW flows transition from integrability to chaos via the formation of chaotic tangles in the aperiodic Lagrangian flow, and establish a direct link between the `lobes' associated with these tangles and intermittency in the observed turbulent dissipation.
This illustrates how utilizing tools from deterministic dynamical systems theory to study convective nonlinearities can explain aspects of intermittent spatiotemporal structure exhibited by the statistics of turbulent fields.
\end{abstract}


\maketitle 

\section{Introduction and background}\label{sec:intro}

Transitionally turbulent flows in both natural and engineered systems often exhibit strong spatiotemporal intermittency, manifesting as observable patterns or structures advected by the flow field. 
Phenomenologically, intermittency refers to the tendency for turbulent behavior to concentrate into structures and/or events which are irregular or sporadic \cite{Frisch1995}.
In neutral fluids, examples include the turbulent ``puffs'' observed in transitional pipe flows \cite{Wygnanski1973,Avila2011}, and the ``plumes'' that detach from viscous boundary layers in thermally-driven convective turbulence \cite{Kadanoff2001,Ahlers2009}.
In magnetically confined fusion plasmas, two key examples include the ``blobs'' or ``filaments'' observed in the scrape-off-layer (SOL) in tokamaks \cite{LaBombard2001,BenAyed2009}, as well as the ``avalanches'' observed in the Dimits shift regime of drift-wave turbulence \cite{beyer2000nondiffusive,mcmillan2009avalanchelike}.

Intermittency poses a major challenge in building quantitative descriptions of turbulence, as the existence of such structures suggests the presence of nontrivial spatiotemporal correlations in the statistics of the turbulence.
Namely, the dynamics of these intermittent structures reflect a complex interplay between their growth from sources of free energy, such as gradients of velocity, temperature, or density, and their resulting enhancement of dissipation and transport of associated quantities, such as momentum, heat, and material.
Thus, building a comprehensive understanding of how turbulent dissipation and transport arise from underlying physical processes in intermittently turbulent flows requires an understanding of the role these structures play in turbulence.

An important example relevant to tokamak plasmas is the Dimits shift regime observed in simulations of drift-wave turbulence with marginally unstable temperature and density gradients \cite{Dimits2000}.
In this regime, interactions between the drift waves and zonal flows organize the turbulence into sporadic radially-localized bursts called ``avalanches'' \cite{beyer2000nondiffusive,Ivanov2020}.
These bursts are observed to organize into nontrivial mesoscale patterns with characteristic length scales \(\rho_s \ll \ell_{meso} \ll L\), where \(\rho_s\) is a characteristic gyroradius and \(L\) is the system size \cite{Jolliet2012,Dif-Pradalier2015a}.
The presence of these mesoscale structures is theorized to underlie the failure of either purely Bohm-like (i.e. system size scaling) or gyro-Bohm-like (i.e. gyroradius scaling) arguments to fully describe observed turbulent transport scalings in tokamaks \cite{Diamond1995,Lin2002,Hahm2018a}.

Another important example relevant to tokamak plasmas is the question of how the radial propagation of structures between the last closed flux surface and open field line regions affects transport in edge and SOL plasmas.
It has been experimentally observed that perturbations in plasma density and temperature in the SOL organize into structures with a blob- or filamentary-like form that is elongated along magnetic field lines and linked to bursts of plasma density and temperature measurements at the tokamak device wall \cite{Kube2020}.
The dynamics of these structures has been suggested to be a form of ``turbulence spreading'' that allows for significant interaction between pedestal and SOL turbulence \cite{Manz2015,Singh2020}.
This plays an important role in setting the SOL heat flux width, a key metric that determines heat flux exhaust requirements in tokamaks.

In this paper, we tackle the issue of intermittency by undertaking a computational and theoretical study of turbulence in the Dimits shift regime in the Hasegawa-Wakatani (HW) equations \cite{majda2018flux,qi2019flux}.
The equations are a model of resistive drift-wave turbulence, and are frequently used as a minimal model for plasma turbulence that includes a physically relevant linear instability along with an \(E \times B\) convective nonlinearity.
Recent works have identified a Dimits shift regime in the flux-balanced Hasegawa-Wakatani (BHW) equations and other related two-dimensional plasma models \cite{St-Onge2017,qi2020dimits,Zhu2020a,Zhu2020b,Ivanov2020}, characterized by strong zonal flows and radially localized bursts of turbulence.
The BHW equations are notable in having an especially robust Dimits shift \cite{qi2020dimits}, which motivates their study here.

The main outcome of this work is to demonstrate that in the Dimits shift regime of BHW turbulence, intermittency in turbulent transport and dissipation can be characterized using tools from deterministic dynamical systems.
The key idea is that the ``fluid parcel relabeling symmetry'', a symmetry generic to neutral fluid and plasma systems \cite{Padhye1996}, can potentially constrain BHW flows to be `nearly-laminar' even when the convective nonlinearity dominates over viscosity.
Drift-wave eigenmodes are shown to possess a certain formal ``near-integrability'' property, extending the results of \cite{cao2023rossby} on Rossby wave turbulence to plasma systems.
The transitionally turbulent Dimits shift regime is then associated with a weakly chaotic regime in the Lagrangian flow arising from the convective nonlinearity.
We show how stable/unstable manifold theory, often used in the study of neutral fluid flows \cite{Balasuriya2016}, in combination with the concept of distinguished hyperbolic trajectories \cite{ide2002distinguished} can be used to understand how intermittency in the Eulerian fields can be directly traced back to the formation of `chaotic tangles' and associated `lobes' in the Lagrangian flow.

This paper is organized as follows:
We begin in Section \ref{sec:bhw} by introducing the Hasegawa-Wakatani equations and its resistive drift-wave instability.
Next, in Section \ref{sec:coherent_vortices} we demonstrate in direct numerical simulations that the Dimits shift regime in BHW is dominated by coherent drift-wave vortices.
Intriguingly, these vortices have a strong linear eigenmode character despite estimates suggesting that nonlinear effects should be comparable with or even dominate linear effects.
This motivates the theoretical discussion in Section \ref{sec:near_integrability}, where the concepts of Generalized Liouville integrability and separatrix splitting are introduced.
It is shown how the drift-wave eigenmodes satisfy a certain formal ``near-integrability'' property, suggesting a route for their survival to large amplitudes.
Furthermore, it is discussed how chaotic tangles associated with hyperbolic trajectories gives a way to understand the transition of flows from integrable to chaotic.
Following this, Section \ref{sec:sums} introduces distinguished hyperbolic trajectories, which are used to accurately realize the theoretical discussions to describe Lagrangian flows in the direct numerical simulations.
In particular, we show how regions of enhanced turbulent mixing and dissipation correlate to the ``lobes'' formed by separatrix splitting, and how this is linked to the survival of the coherent vortices to large amplitude.
Finally, in Section \ref{sec:discussion} we summarize the results and discuss potential applications of the developed techniques to describe the kinematics of coherent structures in the Dimits shift regime and in plasma edge turbulence beyond the model equations studied here.

\section{The Hasegawa-Wakatani equations for plasma edge turbulence} \label{sec:bhw}

\subsection{Fluid formulation for the plasma edge turbulence}

To define the fluid equations, we introduce a doubly-periodic shearless slab geometry in the 2-dimensional domain $(x,y)\in[-L_x/2,L_x/2]\times[-L_y/2,L_y/2]$ as an approximation of the edge of a magnetically confined plasma.
Here $x$ is the coordinate for radial direction and $y$ is the poloidal direction.
A function \(f(x,y)\) on this domain can be decomposed into its zonal mean \(\bar{f}(x)\) and fluctuations about its zonal mean \(\tilde{f}(x,y)\) as
\begin{equation}
    \bar{f} := \frac{1}{L_y} \int f\left(x,y\right)dy, \qquad \tilde{f} = f - \bar{f}.
\end{equation}

The \emph{two-field Hasegawa-Wakatani} (HW) \cite{hasegawa1983plasma} models describe the evolution of the \emph{density fluctuation} $n=n(x,y,t)$ relative to a fixed background density $n_{0}\left(x\right)$, and the \emph{potential vorticity (PV) anomaly} $q=q(x,y,t)=\nabla^{2}\varphi -\tilde{n}-\bar{n}\delta_{s0}$ (with $\delta_{s0}$ the Kronecker delta and $s=0$ or $1$ defining two HW models explained next) by the following equations
\addtocounter{equation}{0}\begin{subequations}\label{eq:plasma_model}
\begin{eqnarray}
\frac{\partial q}{\partial t}+J\left(\varphi,q\right)-\kappa\frac{\partial\varphi}{\partial y} & = & \mu\nabla^2 q,\label{eq:plasma_balance1}\\
\frac{\partial n}{\partial t}+J\left(\varphi,n\right)+\kappa\frac{\partial\varphi}{\partial y} & = & \alpha\left(\tilde{\varphi}-\tilde{n}\right)+\mu\nabla^2 n,\label{eq:plasma_balance2}
\end{eqnarray}
\end{subequations}
where $\varphi$ is the electrostatic potential, and $\mathbf{u} = \left(-\partial_{y}\varphi, \partial_{x}\varphi\right)$ is the $E\times B$ velocity field.
$J(f,g)=\partial_xf\partial_yg-f\partial_xg\partial_y$ is the Jacobian representing flow advection.
The equations are non-dimensionalized so that the ion sound gyroradius is \(\rho_s = 1\).
The constant background density gradient $\kappa\propto -\nabla\ln n_{0}$ is defined by the exponential background density profile near the boundary $n_{0}\left(x\right)$.
The ``adiabaticity parameter'' $\alpha$ determines the degree to which electrons can move rapidly along the magnetic field lines, and is inversely proportional to the resistivity.
On the right hand sides of the equations \eqref{eq:plasma_model}, collisional dissipation effects are introduced with a homogeneous diffusion term $\mu \nabla^2$ on the PV anomaly and on the density.

Notably, the density gradient also gives rise to a quantity called the \textit{potential vorticity} \(q(x,y,t)+\kappa x\), which is a materially conserved scalar invariant in the absence of viscosity.
It satisfies a pure advection-diffusion equation,
\begin{equation} \label{eq:pv_equation}
    \frac{\partial}{\partial t}(q+\kappa x) + \mathbf{u} \cdot \nabla(q+\kappa x) = \mu \nabla^2(q+\kappa x)
\end{equation}

The physical quantities $\varphi$ and $n$ are decomposed into zonal mean states $\overline{\varphi},\overline{n}$ and their fluctuations about the mean $\tilde{\varphi},\tilde{n}$.
In the different Hasegawa-Wakatani models, free-streaming electrons are assumed to have different responses to the zonally-symmetric component of the electrostatic potential, leading to different expressions for the potential vorticity.
We note a key fact that the modified HW (MHW) model \cite{dewar2007zonal} (with $s=0$ in the PV anomaly $q$) includes the zonal mean density $\bar{n}$ in the potential vorticity, while the flux-balanced HW (BHW) model \cite{majda2018flux} (with $s\neq0$) excludes the mean density state in the potential vorticity.
This change ensures that the BHW model has no net radial transport of electrons in the adiabatic limit \(\alpha\to \infty\).

The BHW model offers an improved formulation possessing several desirable properties.
It is shown from rigorous proof and numerical confirmation \cite{majda2018flux,qi2019flux} that in the adiabatic limit, $\alpha\rightarrow\infty$, the BHW model converges to the following one-state equation for the PV anomaly
\begin{equation}
\frac{\partial q}{\partial t}+J\left(\varphi,q\right)-\kappa\frac{\partial\varphi}{\partial y}=\mu\Delta q,\quad q=\nabla^{2}\varphi-\tilde{\varphi},\label{eq:plasma_onelayer}
\end{equation}
which is called the \emph{modified Hasegawa-Mima} model \cite{itoh2006physics}.
It was shown in \cite{qi2020dimits} that these changes result in an especially robust Dimits shift regime for the BHW model, which we exploit here to study the dynamics of drift waves interacting with strong zonal flows.

\subsection{Linear instability for resistive drift waves}

In the Hasegawa-Wakatani equations, the drift wave instability is due to the non-adiabatic electron response at finite resistivity \(\alpha < \infty\).
We start with single wavenumber fluctuation modes $\left(\tilde{q},\tilde{n}\right)$
with $k_{y}\ne0$ in the following form (the subscripts $k$ for the
single mode variables are neglected for simplicity)
\begin{equation}
\tilde{q}=\hat{q}e^{i\left(\mathbf{k\cdot x}-\omega t\right)},\;\tilde{n}=\hat{n}e^{i\left(\mathbf{k\cdot x}-\omega t\right)},\quad\tilde{\varphi}=\hat{\varphi}e^{i\left(\mathbf{k\cdot x}-\omega t\right)}=-k^{-2}\left(\hat{q}+\hat{n}\right)e^{i\left(\mathbf{k\cdot x}-\omega t\right)}.\label{eq:modes}
\end{equation}
Above $\omega=\omega\left(\mathbf{k}\right)$ is the wave frequency
for each corresponding wavenumber. The single mode potential function
is recovered from the fluctuation state relation $\tilde{q}=\nabla^{2}\tilde{\varphi}-\tilde{n}$.
Substituting the above single-mode expressions \eqref{eq:modes} into
the original equation \eqref{eq:plasma_model}, the nonlinear coupling
terms, $J(\varphi, q)$ and $J(\varphi, n)$,
vanish, resulting in a linearized system. The coefficients
$\left(\hat{q},\hat{n}\right)$ then form a $2\times2$ linear subsystem
for each wavenumber $\mathbf{k}$.

Considering the case \(\mu=0\), the pair of eigenvalues $\omega^{\pm}=\pm\varpi+i\gamma^{\pm}$
give the two branches of eigendirections for the two state pair $\left(\hat{q}^{\pm},\hat{n}^{\pm}\right)$
and the corresponding potential function $\hat{\varphi}^{\pm}$ as
\begin{equation}
\hat{n}^{\pm}=\left(\kappa^{-1}k_{y}^{-1}k^{2}\omega^{\pm}-1\right)\hat{q}^{\pm},\;\hat{\varphi}^{\pm}=-\kappa^{-1}k_{y}^{-1}\omega^{\pm}\hat{q}^{\pm}.\label{eq:lin_rela}
\end{equation}
The unstable direction with $\gamma^{+}>0$ implies exponential growth in drift wave amplitude
along this eigendirection $(\hat{q}^{+},\hat{n}^{+})$ from the linear drift instability; the stable
direction $\gamma^{-}<0$ implies exponential decay correspondingly along the other eigendirection $(\hat{q}^{-},\hat{n}^{-})$.
The entirely explicit formulas for the two branches of the eigenvalues
$\omega^{+}=\varpi+i\gamma^{+}$ and $\omega^{-}=-\varpi+i\gamma^{-}$
can be computed as
\begin{equation}\label{eq:growth}
\begin{aligned}\varpi= & \frac{\mathrm{sgn}\left(k_{y}\right)\alpha}{2\sqrt{2}}\left(1+k^{-2}\right)\left(\sqrt{1+16\kappa_*^{2}}-1\right)^{\frac{1}{2}},\\
\gamma^{\pm}= & \pm\frac{\alpha}{2}\left(1+k^{-2}\right)\left[\frac{1}{\sqrt{2}}\left(\sqrt{1+16\kappa_*^{2}}+1\right)^{\frac{1}{2}}\mp1\right],
\end{aligned}
\end{equation}
with the wave and growth coefficients majorly dependent on the parameter ratio $\kappa_* :=\frac{\kappa}{\alpha}\frac{k_{y}k^{2}}{\left(1+k^{2}\right)^{2}}$.
Thus, the features of the linear instability depend primarily on the ratio $\kappa/\alpha$.

\subsection{Dimits shift regime with intermittent flux transport}
The \emph{Dimits shift regime} of the BHW model refers to a transitional regime of turbulence where \(\kappa/\alpha\) is large enough to sustain a strong zonal flow regime driven by the drift wave instability, but small enough that the zonal flows are able to significantly quench the instabilities before they realize a fully turbulent state \cite{qi2020dimits}.
Before the Dimits shift regime $\kappa/\alpha\ll1$ when all modes are stable due to dissipation, the flow remains laminar with no turbulence.
As the ratio $\kappa/\alpha$ increases, the flow enters the Dimits regime where richer structures are gradually induced due to the excited drift waves from the positive growth rate in \eqref{eq:growth}.
The excited drift waves transfer much of their energy to zonal flows and fluctuation modes through nonlinear interactions, leading to dynamics which are dominated by long-lived ``nearly-laminar'' zonal flows.
Typically, intermittent radial fluxes will gradually develop in time as the secondary instability saturates.

\begin{figure*}

\centering
\includegraphics[scale=1.]{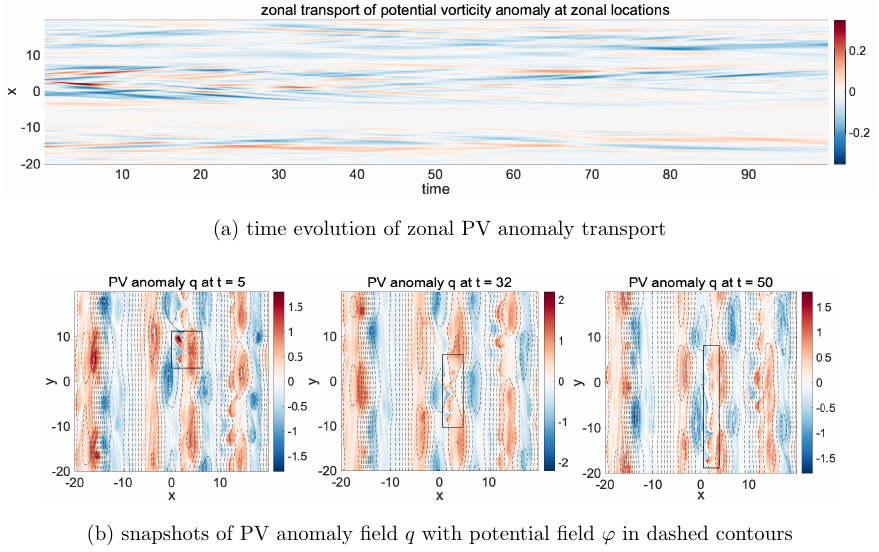}

\caption{(a) Zonal PV anomaly transport $\overline{uq}$ (with $u$ the zonal velocity and $q$ the PV anomaly) together with (b) several typical snapshots
of the PV anomaly field $q$ at different time instants in the Dimits regime. A typical sheared vortex by the zonal flow field is identified in the black box.\label{fig:Zonal-transport} }
\end{figure*}

Examination of the Dimits shift regime reveals a rich variety of both chaotic and regular flow patterns.
As a typical illustration of flow structures in the Dimits regime, we plot the flow transport and interactions with coherent structures generated in the direct numerical simulation solution in weakly chaotic regime in Fig.~\ref{fig:Zonal-transport}.
We also plot the averaged zonal PV anomaly transport $\overline{uq}=L_{y}^{-1}\int uq dy$.
As established in earlier work \cite{qi2019flux,qi2020dimits}, intermittent bursts of zonal fluxes are observed as the flow evolves in time, indicating potential vorticity transport across the zonal flow barriers.
For example, the zonal transport around \(x \approx 0\) is strong at \(t \approx 0\), but this zonal transport then decays as the simulation continues.

To look at this transport event more closely, in the second row of Fig.~\ref{fig:Zonal-transport} we also plot the snapshots of PV anomaly field.
As indicated in the black box, the concentrated small vortex in the region $x\in\left[0,5\right]$ is gradually sheared by the zonal flows and coherent vortices in the neighborhood in $x\in\left[-2,8\right]$ moving in opposite directions along $y$.
Notably, this transport event does not appear to strongly perturb the coherent vortices, and does not become a radially-extended avalanche.
This indicates the typical flow transporting features in the weakly turbulent regime of BHW.
In the next sections, we will focus on this particular transitional regime of turbulence and try to elucidate the dynamics of the zonal flows and coherent vortices, and how they organize turbulent transport and dissipation.

\section{Identifying coherent drift-wave vortices} \label{sec:coherent_vortices}

In this section, we investigate turbulent structures in the Dimits shift regime as the ratio $\kappa/\alpha$ stays in a relatively small value.
In particular, we characterize the radially localized vortices which are observed in the Dimits shift regime, and discuss how they show a significant linear character despite the nonlinear convection term being dominant.

\subsection{Data-driven identification of drift-wave structures in simulations}\label{subsec:pod}

\begin{figure*}
    \centering
    \includegraphics[scale=1.]{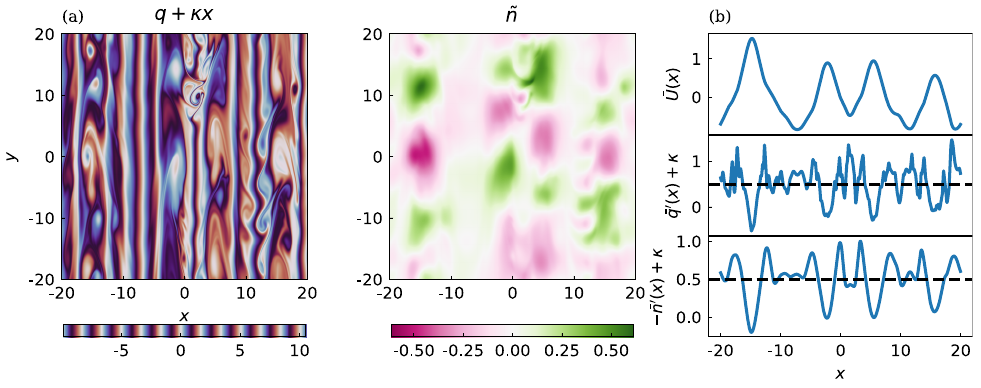}
    
    \caption{\label{fig:dns_snapshot} (a) DNS snapshot of the potential vorticity \(q+\kappa x\) and non-zonal component of the density \(\tilde{n}\). (b) Zonal averages from the snapshot of the zonal velocity, PV gradient, and density gradient.}
\end{figure*}

We start with analyzing direct numerical simulations (DNS) of turbulence in the BHW model.
Parameters were taken so that the turbulence would be in the Dimits shift regime of BHW turbulence identified in \cite{qi2019flux}. A standard pseudo-spectral scheme with $N_x$ modes in $x$ and $N_y$ modes in $y$ is used for spatial discretization with a fourth-order Runge-Kutta scheme for accurate time integration.
See Table \ref{tab:dns_params} for a list of parameters used in this study, and see Fig.~\ref{fig:dns_snapshot} for an example snapshot.

\begin{table}
\caption{\label{tab:dns_params} Table of DNS parameters used in this study}
\begin{tabular}{lcc}
\textbf{Name} & \textbf{Symbol} & \textbf{Value} \\
\hline
Resolution & \(N_x \times N_y\) & \(1024 \times 1024\) \\
Box size & \(L_x \times L_y\) & \(40 \times 40\) \\
Density gradient & \(\kappa\) & \(0.5\) \\
Adiabaticity & \(\alpha\) & \(1.0\) \\
Diffusivity & \(\mu\) & \(1\times 10^{-3}\)
\end{tabular}
\end{table}

From Fig.~\ref{fig:dns_snapshot}, several observations can be made about the turbulence.
Significant perturbations to both the density and PV develop in both zonal and non-zonal components of the flow.
Corrugations develop in the zonally-averaged PV and density profiles, with regions of mixing strong enough to completely flatten the background PV gradient \(\kappa\), i.e. \(k_Z \bar{q} \sim k_Z \bar{n} \sim \kappa\) where \(k_Z\) is a characteristic zonal wavenumber.
This leads to a discrete set of banded jets, similar to an \(E \times B\) staircase \cite{Dif-Pradalier2010a}, which alternate between propagation in the \(+y\) ion diamagnetic drift direction and \(-y\) electron diamagnetic drift direction.
In turn, the amplitude of non-zonal perturbations of both PV and density can be just as large as the zonal corrugations, i.e. \(\tilde{n}/\bar{n} \sim \tilde{q}/\bar{q} \sim O(1)\).

Since the viscosity \(\mu\) is small, structure in the PV \(q+\kappa x\) tends to reflect the structure of the underlying advective flow.
Looking at the zonal flow profile, the regions of strongest shear appear to suppress turbulent mixing, leading to regions where the PV and density gradients are maintained above the background value of \(\kappa\).
PV transport is also suppressed in the tips of \(-y\) directed zonal jets, suggesting the presence of non-twist transport barriers that can form at zero-crossings of the \(E \times B\) shear \cite{Del-Castillo-Negrete1996}.
Visual inspection of Fig.~\ref{fig:dns_snapshot} also shows coherent vortex structures in the PV, with the largest ones spanning about \(\sim5\) gyroradii in radial extent and \(\sim 10\) gyroradii in poloidal extent.
The large vortices are typically localized to \(+y\) directed zonal jets and propagate with a poloidal speed similar to the zonal flows.
The vortices are also associated with large coherent blob-like disturbances to the plasma density field.

To quantify how much of the kinetic and potential energy is stored in coherent structures, we use the proper orthogonal decomposition (POD).
POD has been applied in the past to turbulent flows \cite{Berkooz1993,Smith2005}, and have also been applied specifically to plasma turbulence \cite{Futatani2009,Hatch2011a}.
In order for the POD to have physical meaning, it is important to choose the right observables and inner product for the snapshots.
Here, we use the vorticity and density fields as observables for a sequence of DNS snapshots corresdponing to \(s=200\) consecutive timesteps with \(\Delta t = 0.5\) (that is, a time window of size \(T=100\)).
We then adopt an energy inner product, which allows the decomposition of these DNS snapshots into orthogonal modes, i.e. \(q(x,y,t) = \sum_k a_k(t) q_k(x,y)\) and \(n(x,y,t) = \sum_k a_k(t) n_k(x,y)\), based on their contribution to the total energy of the system.
See Appendix~\ref{app:pod} for the details of the POD analysis.

\begin{figure*}
    \centering
    \includegraphics[width=\linewidth]{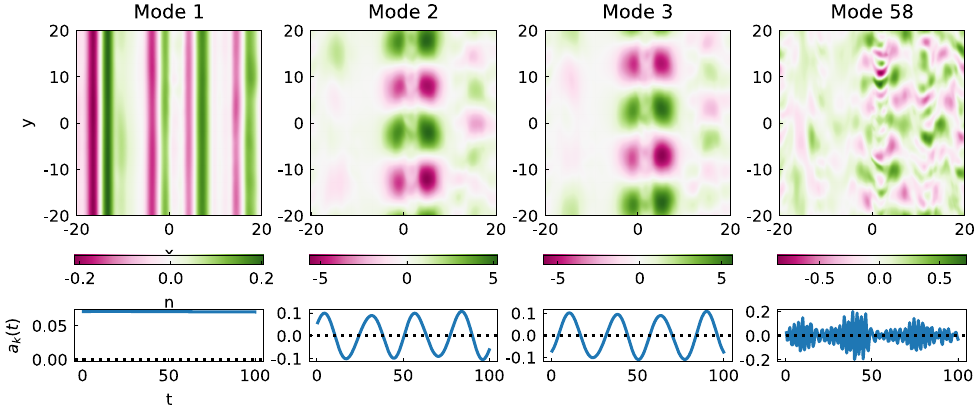}
    \caption{\label{fig:pod_example} Examples of POD modes, showing (top) the density component of the POD mode \(n_k(x,y)\), and (bottom) the corresponding POD mode amplitudes \(a_k(t)\).}
\end{figure*}

Examples of POD modes computed from the DNS snapshots are shown in Fig. \ref{fig:pod_example}.
Mode 1 is the most energectic POD mode, and contains both a nearly zonally-constant zonal flow and zonal density component.
These next few modes exhibit strongly structured behavior in both time and space, with sinusoidal behavior time \(t\) and in the zonal direction \(y\), and a coherent envelope structure in the radial direction \(x\).
For example, mode 2 is dominated by two bands of alternating density perturbations centered in the \(+y\) directed zonal jets at \(x\approx -2.5\) and \(x \approx 5\).
Mode 2 and mode 3 form a sine/cosine pair that can be interpreted as a single traveling wave which travels in the \(+y\) direction through the trigonometric identity \(\cos(\omega t - k y) = \cos(\omega t) \cos(k y) + \sin(\omega t) \sin(k y)\).
Meanwhile, higher modes are increasingly incoherent in both time and space.
For example, while mode 58 has some recognizable features such as the disturbances propagating in \(-y\) centered around \(x \approx 2\), these features do not have a cleanly isolated spatial signal nor a coherent time signature.

\begin{figure}
    \centering
    \includegraphics[scale=1.]{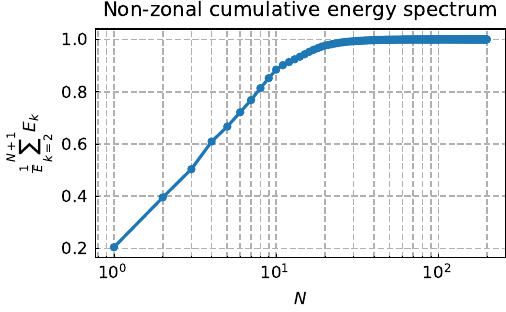}
    \caption{\label{fig:pod_spectrum} Cumulative energy spectrum for the non-zonal POD modes.}
\end{figure}

Approximately 93\% of the energy is stored in mode 1, so the vast majority of the energy is contained in the zonal density and zonal flow components.
For the non-zonal energy, a very large fraction is captured by a small number of modes, as only 10 additional POD modes are needed to capture 90\% of the non-zonal energy.
This is illustrated in the cumulative energy spectrum for non-zonal modes in Fig. \ref{fig:pod_spectrum}.
Most of these modes have coherent spatiotemporal structure, like the shown mode 2 and mode 3.
Thus, the majority of the turbulent energy is contained in modes with significant spatiotemporal coherence.

\subsection{Drift-wave eigenmode character of coherent structures} \label{sec:eigenmodes}

Next, we establish that the coherent structures identified in the previous section are in fact large-amplitude drift-wave eigenmode structures.
Given the large-amplitude corrugations that form in the PV and density profiles, it makes sense to linearize around the zonally and temporally-averaged PV and density profiles.
Taking \(q = \bar{q}(x) + \hat{q}(x) e^{i (k_y y - \omega t)}\), \(n = \bar{n}(x) + \hat{n}(x) e^{i (k_y y - \omega t)}\), and with a background zonal flow profile \(U(x)\), an equation for the eigenvalues and eigenfunctions can be derived:
\begin{subequations}\label{eq:eig}
    \begin{align}
        \omega \hat{q}(x) &= k_y U(x) \hat{q}(x) - k_y (\bar{q}^{\prime}(x) + \kappa) \hat{\varphi}(x) + i\mu (\partial_x ^2 - k_y^2) \hat{q}(x) \label{eq:eig_q}, \\
	\omega \hat{n}(x) &= k_y U(x) \hat{n}(x) - k_y (\bar{n}^{\prime}(x) - \kappa) \hat{\varphi}(x) + i\alpha(\hat{\varphi}(x) - \hat{n}(x)) + i \mu (\partial_x^2 - k_y^2) \hat{n}(x), \\
	\hat{\varphi}(x) &= (\partial_x^2 - k_y^2)^{-1} \left[\hat{q}(x) + \hat{n}(x)\right].
    \end{align}
\end{subequations}

In the following, we will refer to solutions of \eqref{eq:eig} as the \emph{eigenmodes}, while we refer to the solutions to the linear problem in the absence of a non-trivial background profile given by \eqref{eq:lin_rela} and \eqref{eq:growth} as the \emph{dispersion relation modes}.
We will write \(\omega_{eig} = \varpi_{eig} + i \gamma_{eig}\) for the eigenfrequencies and \(\omega_{disp}\) for the dispersion relation frequencies when it is necessary to disambiguate.
Unlike the dispersion relation modes which have the structure of Fourier modes, the eigenmodes have a non-trivial standing wave structure in the radial direction \(x\).
The eigenmodes propagate in the zonal direction \(y\) with a phase velocity \(u_{ph} = \varpi_{eig} / k_y\).

\begin{figure*}
    \centering
    \includegraphics[width=\linewidth]{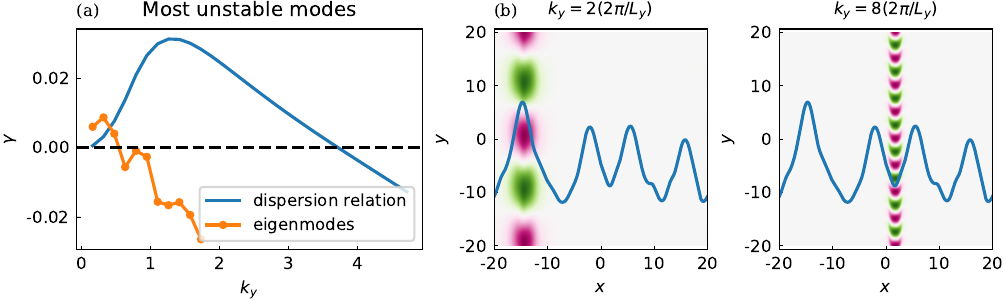}
    \caption{(a) Plot of the growth rate \(\gamma\) for the most unstable mode at each \(k_y\) for the dispersion relation modes (blue curve) versus the eigenmodes (orange curve with dots). (b) A pair of plots showing the density \(\tilde{n}\) for the most unstable eigenmodes at two different values of \(k_y\), with the zonal flow profile \(U(x)\) overplotted.}
    \label{fig:eigenmodes_vs_dispersion}
\end{figure*}

One key feature of the eigenmodes is that they are usually radially localized, taking on a non-zero \(k_x \sim k_y\) and inducing a vortex-like flow.
By contrast, the most unstable dispersion relation modes tend to induce streamer-like flows with \(k_y \sim 1/\rho_s = 1\) and \(k_x = 0\), so this radial localization tends to severely limit the growth rate of the eigenmodes compared to the dispersion relation modes.
This effect is illustrated in Fig.~\ref{fig:eigenmodes_vs_dispersion}, which shows how the growth rates and radial structure of the eigenmodes differ from those of the dispersion relation modes.

Note that the gradient of the background PV \(\bar{q}^{\prime}(x) + \kappa\) is not sign-definite, so the zonal flow profile would not satisfy the Rayleigh-Kuo criterion for stability against Kelvin-Helmholtz type shear instabilities.
However, as discussed in \cite{Zhu2018b,Zhu2020a} since the dominant zonal flow wavenumber does not exceed the inverse ion sound gyroradius, the instabilities here can be understood as modified primary (i.e. resistive drift-wave-like) instabilities, rather than as shear-flow instabilities.
Their radially-localized structure can then be understood through the structure of the Drifton phase space \cite{Zhu2018a}, or through the harmonic oscillator approximation \cite{Zhu2020a}.
We observe a similar phenomenology to the latter study, with \(k_y \sim 1/L_y\) trapped modes in the \(+y\) directed zonal jets, and higher \(k_y \sim 1/\rho_s\) runaway modes in the \(-y\) directed zonal jets.

\begin{figure*}
    \centering
    \includegraphics{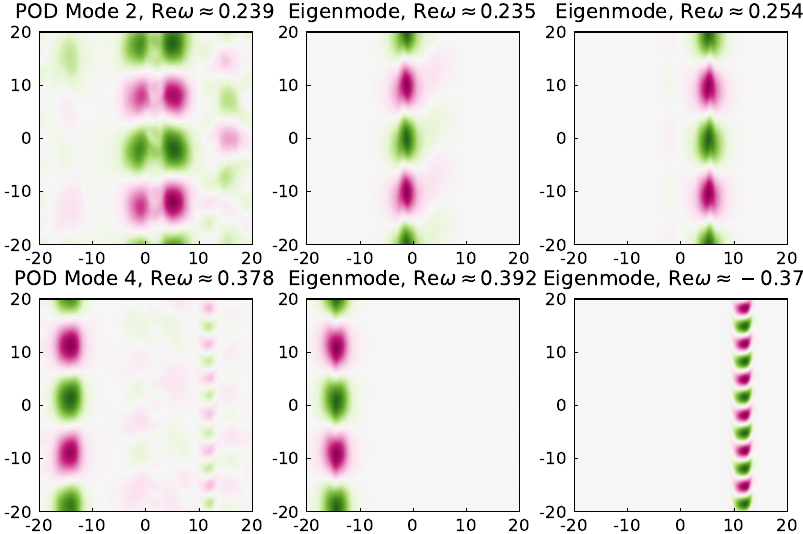}
    \caption{Plot of density field \(n_k(x,y)\) for POD modes (left) and correpsonding density fields for eigenmodes (center and right).}
    \label{fig:pod_comparison}
\end{figure*}

We can compare the structure of the computed eigenmodes to the POD modes computed in the previous section.
We find that the coherent POD modes tended to be the combination of several eigenmodes of similar eigenfrequency.
This is illustrated in Fig.~\ref{fig:pod_comparison} for POD mode 2 and POD mode 4.
In POD mode 2, the density perturbations near the center of the domain are seen to correspond to two different eigenmodes with similar frequency.
Note that the phase correlation of the eigenmodes appears to be a coincidence, as inspection of the full flow field shows that the coherent vortices corresponding to the two different eigenmodes drift with respect to each other over time.
Meanwhile in POD mode 4, the density perturbations centered near \(x \approx -15\) correspond to one eigenmode.
Furthermore, there is a faint band of coherent perturbations near \(x \approx 11\), which also corresponds to an eigenmode, although it has a negative rather than positive frequency.
In both cases, the overall structure of the primary perturbations are captured, although there is some difference in the exact shapes of the perturbations.

\begin{figure*}
    \centering
    \includegraphics[width=\linewidth]{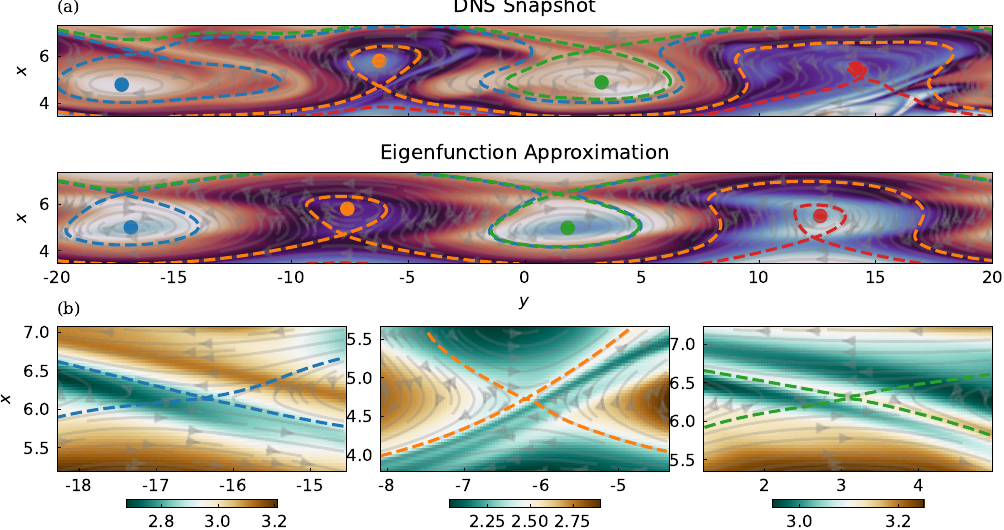}
    \caption{(a) Comparison of PV \(q + \kappa x\) for the DNS snapshots (top) and a selection of three eigenfunctions (bottom). Streamlines are also shown for the frozen flow in a frame of reference co-moving with the vortices. The vortex separatixes are shown with dashed lines, and corresponding O-points marked in color. (b) A series of zoomed-in pictures near the X-points, which show the detail how perturbations in the PV field align with the vortex separatrixes.}
    \label{fig:eig_dns}
\end{figure*}

Beyond the POD modes, we observe that the eigenmodes are also able to capture key ``coarse-grained'' features of the coherent vortices in the DNS.
A plot of the sum of three selected eigenmodes compared to the DNS is shown in Fig.~\ref{fig:eig_dns}, focusing on the \(+y\) directed zonal band centered around \(x\approx 5\).
The eigenfunctions are able to reproduce the rough location of instantaneous stagnation points in the reference frame comoving with the coherent vortices, that is points \((x,y)\) at some fixed time \(t\) where \(\mathbf{u}(x,y,t) - (u_{ph}, 0) = 0\).
Since the flow is incompressible, these stagnation points take the form of either O-points or X-points in the streamfunction \(\varphi - u_{ph} x\).
The O-points correspond to the center of the coherent vortices.
Meanwhile, the streamlines connected to the X-point define separatices, which separate the vortical flow in the vortexes from the topologically distinct zonal flow outside of the vortices.

We remark here that the topology of the coherent vortex flow resembles the structure of homoclinic island chains observed in non-twist Hamiltonian flows \cite{DelCastilloNegrete1993,Del-Castillo-Negrete1996,Del-Castillo-Negrete2000}.
This resemblance holds despite the differing physical origin of the wave instability, as the aforementioned studies considered drift-Rossby waves driven unstable by flow shear, whereas here the main instability mechanism arises from resistivity in the presence of a density gradient.
The flow topology also bears resemblance to the sinuous jet structure observed in simulations of two-dimensional Couette flow \cite{Falkovich2018}.
The vortices also appear to resemble the monopolar vortices which are approximate exact solutions to the Hasegawa-Mima equations studied in Nycander \cite{Nycander1988}.
Note that the eigenfunctions do not correctly capture the fine-scale details of the PV field, especially the filamentary structure of the PV near the vortex boundaries.
However, the filamentary structures do tend to align with the unstable manifolds of the frozen flow, suggesting their dynamical relevance.

\begin{figure}
    \centering
    \includegraphics{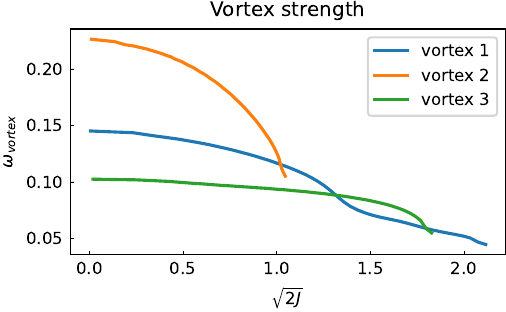}
    \caption{Strength of the three largest vortices in Fig.~\ref{fig:eig_dns} as measured by \(\omega_{vortex}\), the frequency of the vortex rotation.}
    \label{fig:vortex_strength}
\end{figure}

Given that the eigenfunctions seem to capture much of the qualitative behavior of the coherent vortices, it may be tempting to conclude that the vortices are simply linear structures.
One way to provide a quantitative measure of the strength or level of nonlinearity of the vortices is to compute the vortex rotation rate \(\omega_{vortex}\), defined as the rate at which a particle following the flow executes motion around the vortex.
\(\omega_{vortex}\) is the principal nonlinear timescale associated with the convective nonlinearity of the vortex flow.
It can be computed from the relation \(\omega_{vortex} = d\Psi(J)/dJ\), where \(\Psi(J)\) is the value of the streamfunction in the comoving frame on a closed contour of the streamfunction, and \(J\) is the action variable equal to the area of enclosed by the closed streamfunction contour divided by \(2 \pi\).
If a fluid parcel travels at a typical speed \(U\) in a circular vortex of radius \(L \sim \sqrt{2 J}\), then \(2 \pi L / U = 2 \pi / \omega_{vortex}\), which means \(\omega_{vortex} \sim U/L\).
This measure of vortex strength is plotted in Fig.~\ref{fig:vortex_strength} for the three primary vortices observed in the zonal band plotted in Fig.~\ref{fig:eigenmodes_vs_dispersion}.

Using the characteristic lengthscale \(L=2\) for the vortices, we can compute the Reynolds number \(Re \sim U L / \mu = \omega_{vortex} / (\mu / L^2) \approx 580\).
Thus, PV transport by the convective nonlinearity dominates over viscous transport in the vortex.
For density transport, we observe that the vortices are roughly adiabatic, with \(\omega_{vortex} / \alpha \approx 0.15\).
In the adiabatic limit where \(\alpha \to \infty\), the dynamics of the system becomes purely determined by the PV transport, so from these two measures we conclude that the vortex dynamics appears to be dominated by the nonlinear convective PV transport.

Next, it may be possible to use linear response theory to describe the trajectory of fluid parcels in the vortex flow even if the convective nonlinearity is large or dominant.
This is a standard assumption behind quasilinear theory \cite{Diamond2010,Krommes2015}.
Taking the assumption that nonlinear convection should balance the linear growth rate of the eigenmodes, the typical autocorrelation time $\gamma_{ac}$ of the flow experienced by a fluid parcel can be estimated by \(\gamma_{ac} \sim \gamma_{eig}\), where $\gamma_{eig}$ is defined below \eqref{eq:eig}.
From this we can estimate the Kubo number \(Ku := \omega_{vortex} / \gamma_{ac} \approx 7.3\).
Thus, we find that the flow here is in the vortex trapping regime, where the usage of linear response theory is unjustified since the coherent vortices make strongly nonlinear deflections to the fluid parcel trajectories.

For the spatially localized eigenfunctions observed here, one further comparison between nonlinear and linear timescales can be made between the typical velocity fluctuation around the zonal flow centered at the eigenfunctions \(\delta u_{rms}(x_{eig}) := \sqrt{\left\langle |u(x_{eig},y,t) - U(x_{eig})|^2 \right\rangle}\), and the relative phase velocity of the eigenmodes relative to the zonal flows, \(\delta u_{ph} := u_{ph} - U(x_{eig})\).
If \(|\delta u_{ph}| \gg \delta u_{rms}\), then in a frame of reference following the zonal flows, fluctuations in the field undergo linear oscillations many times before they are scattered by nonlinear convection.
Meanwhile if \(\delta u_{rms} \gg |\delta u_{ph}|\), fluctuations would be scattered by nonlinear convection before they take on any wave-like quality.
For the eigenfunctions used in Fig.~\ref{fig:eig_dns}, the PV and density envelopes peak near \(x_{eig} \approx 5.4\), with \(\delta u_{rms}(x_{eig}) \approx 0.054\) and \(|\delta u_{ph}| \approx 0.1\) for all three eigenfunctions.
Thus, we estimate that the observed vortices are ambiguously nonlinear, with \(|\delta u_{rms} / \delta u_{ph}| \approx 0.5\).

\subsection{Spatially intermittent patterning of turbulence}

Now, we move to the description of the regions of turbulence that coexist with the coherent vortices, and discuss how they display a form of spatially intermittent patterning.
Past studies have connected the properties of the Lagrangian flow in Hasegawa-Wakantani to intermittency, see e.g. \cite{Kadoch2010,Kadoch2022}.
The concepts of spatial coherence and intermittency are closely related, as the turbulent dissipation and mixing tends to spatially concentrate outside of the coherent structures.

Based on the advection-diffusion equation \eqref{eq:pv_equation} for the PV, we consider two ways to measure the turbulent dissipation and mixing.
The first is the magnitude of the PV gradient \(|\nabla(q+\kappa x)|\), which is related to the flux of potential vorticity due to viscosity \(\mu \nabla (q+\kappa x)\).
Turbulent mixing of the PV will typically lead to the formation of small-scale structure which needs to be dissipated by viscosity, so the viscous flux magnitude is typically far above the background value of \(\mu\kappa\).
The second is the radial PV flux divergence \(u \partial_x (q + \kappa x)\).
This gives a more direct measure of PV advection, although only chaotic or stochastic advection would typically be expected to lead to irreversible mixing.

To define intermittency in the DNS, we consider a theoretical baseline given by the \textit{random phase (RP) approximation}, a standard assumption used in weak wave turbulence \cite{Nazarenko2011,Krommes2015}.
For a given snapshot of the DNS, we define an analogue random phase (RP) field by
\begin{subequations}
    \begin{align}
        q_{RP}(x,y) &= \sum_{\mathbf{k}}\left[|\hat{q}^{+}_{\mathbf{k}}|e^{i\mathbf{k} \cdot \mathbf{x}+i\theta^{+}_{\mathbf{k}}} + |\hat{q}^{-}_{\mathbf{k}}|e^{i\mathbf{k} \cdot \mathbf{x}+i\theta^{-}_{\mathbf{k}}}\right], \\
        n_{RP}(x,y) &= \sum_{\mathbf{k}}\left[|\hat{n}^{+}_{\mathbf{k}}|e^{i\mathbf{k} \cdot \mathbf{x}+i\theta^{+}_{\mathbf{k}}} + |\hat{n}^{-}_{\mathbf{k}}|e^{i\mathbf{k} \cdot \mathbf{x}+i\theta^{-}_{\mathbf{k}}}\right].
    \end{align}
\end{subequations}
Here, \((\hat{q}^{\pm}_{\mathbf{k}}, \hat{n}^{\pm}_{\mathbf{k}})\) are the complex amplitudes of the dispersion relation modes \eqref{eq:modes} taken from the DNS.
We take the absolute value so the phase information is lost, then multiply by a random phase \(\theta^{\pm}_{\mathbf{k}}\) that is uniformly distributed on \([0,2\pi]\) and independent between different \(\mathbf{k}\), except as necessary to maintain the real-valuedness of the fields.
This models a bath of dispersion relation modes, depending only on the ``equilibrium'' gradient \(\kappa\).
This bath has an identical spectrum of modes to the DNS, but has lost all spatial information carried in the phase relationship between different \(\mathbf{k}\) modes.

\begin{figure*}
    \centering
    \includegraphics[width=\linewidth]{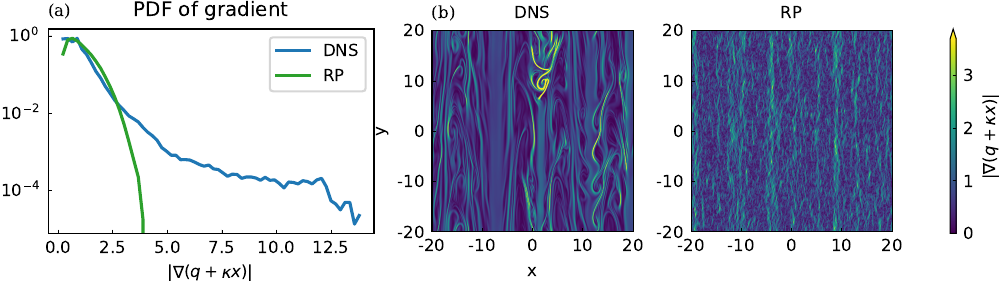}
    \caption{(a) Histogram showing the PDF of the magnitude of the PV gradient \(|\nabla(q+\kappa x)|\) for one time snapshot from the DNS, and with random phases (RP). (b) Spatial distribution of the magnitude of the PV gradient. Note that large values in the DNS field are clipped to show contrast.}
    \label{fig:intermittency}
\end{figure*}

\begin{figure*}
    \centering
    \includegraphics[width=\linewidth]{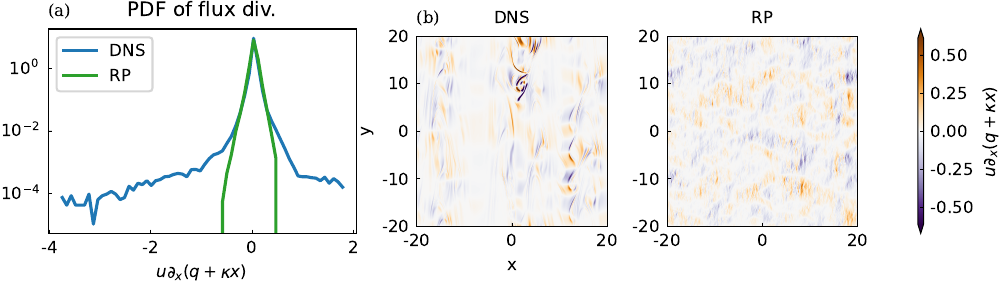}
    \caption{(a) Histogram showing the PDF of the radial PV flux divergence \(u \partial_x (q+\kappa x)\) for one time snapshot from the DNS, and with random phases (RP). (b) Spatial distribution of the radial PV flux divergence. Note that large values in the DNS field are clipped to show contrast.}
    \label{fig:intermittency_flux}
\end{figure*}

As a result of this loss of spatial information, the RP fields tend to be less intermittent, in that the turbulent dissipation and mixing tends to be less spatially concentrated.
We illustrate this for the DNS snapshot in Fig.~\ref{fig:dns_snapshot}.
Consider first the magnitude of the PV gradient, shown in Fig.~\ref{fig:intermittency}.
The total turbulent dissipation, measured by the integral \(\int \mu | \nabla (q+\kappa x) | ^2 d\mathbf{x}\), differs by less than 1\% between the DNS and RP fields.
However, in the PDF of the PV gradient, the original DNS field has a much heavier tail than the RP field.
Additionally, the turbulent dissipation has a much more defined filamentary structure in the original DNS field that correlates with the bands of vortices.
A similar conclusion holds for the radial PV flux divergence, shown in Fig.~\ref{fig:intermittency_flux}.
Again, the tails of the PDF are much heavier for the DNS field, and the flux exhibits a much more defined filamentary structure.
Thus, the observed DNS fields tend to be much more intermittent in these measures of turbulence than expected for the corresponding RP fields.

In summary, eigenfunctions appear to be able to capture many qualitative features of the flow that correlate with intermittency in turbulent dissipation.
However, nearly every measure of nonlinearity shows that the flow is either ambiguously nonlinear or dominated by the nonlinearity.
This leads to the important question that deserves more explanation: why is it that drift-wave eigenmodes can survive to large amplitude while maintaining linear coherence?
Furthermore, what role do the vortex separatrices play in creating the filamentary structure in the turbulent intermittency?
In the next sections, we show that the answers to these questions are related through the concept of near-integrability and weak chaos in Lagrangian flows.

\section{General theory of near-integrability of drift-wave Lagrangian flows} \label{sec:near_integrability}

To motivate this section, we observe that the convective nonlinearity in the Hasegawa-Wakatani equations takes the form of an advection equation by a time-dependent Hamiltonian \(-\varphi(x,y,t)\).
This advection has associated characteristic equations which determine the motion of tracer particles in the flow, also known as the Lagrangian flow equations which determine the Lagrangian trajectories.
These are given by the set of ordinary differential equations
\begin{subequations} \label{eq:hamiltonian}
    \begin{eqnarray}
    \frac{dX(t)}{dt} = \{x,-\varphi\} = u(X(t),Y(t),t), \\
    \frac{dY(t)}{dt} = \{y,-\varphi\} = v(X(t),Y(t),t),
    \end{eqnarray}
\end{subequations}
where $\mathbf{u}=(-\partial_y \varphi,\partial_x \varphi)$ is the flow field and $\{f,g\}$ is the Poisson bracket.
Referring to equation \eqref{eq:plasma_balance1}, the PV \(q(x,y,t) + \kappa x\) can be shown to serve as an approximate integral of motion for this flow, with
\begin{equation}\label{eq:hamil_flow}
    \frac{d}{dt}\left[q(X(t),Y(t),t) + \kappa X(t)\right] = O(\mu)
\end{equation}
along the Lagrangian trajectories \(\mathbf{X}(t)=(X(t),Y(t))\) given by \eqref{eq:hamiltonian}.
If \(\mu = 0\), then the PV would be an exact integral of motion.

In the following, we show how the existence of this approximate integral of motion relates to a certain ``near-integrability'' property of the drift-wave eigenmodes, suggesting that the Lagrangian flow induced by the zonal flows plus drift waves is primarily \textit{laminar}.
Furthermore, it is discussed how integrability transitions to chaos through the formation of chaotic tangles, providing a way to characterize the development of spatially localized regions of chaotic mixing in the weakly turbulent regime.

In this section, we focus on developing these ideas in the context of fairly generic dynamical systems, giving ways to generalize the results here beyond the Hamiltonian flows relevant to the Hasegawa-Wakatani equations.
We first introduce the notion of \textit{Generalized Liouville integrability}, which provides a framework for integrability that can apply to flows which are not necessarily Hamiltonian.
The connection between symmetry and integrability is emphasized, showing how the dynamical properties of the flow can be understood to originate from physical properties of the fluid system.
Using this framework, we derive the near-integrability property of the drift-wave eigenmodes.
We follow this with a brief overview of stable/unstable manifold theory, which provides a systematic way to understand how integrable systems transition to chaos.
This lays the groundwork for section \ref{sec:sums}, where we show how to apply these theories to characterize spatial intermittency in the observed DNS flows.

\subsection{Symmetries and Generalized Liouville integrability}

Here, we review the notion of Generalized Liouville integrability, following \cite{Zung2016}.
Recall that in Hamiltonian systems, Liouville integrability says that when a system has enough conserved quantities that satisfy certain conditions, trajectories of the system will undergo quasiperiodic motion confined to invariant tori that foliate the phase space.
Physically, this implies that the phase space flow of an integrable Hamiltonian system is in fact `laminar', with the `laminae' consisting of the invariant tori.
The key idea is that Generalized Liouville integrability provides a systematic extension of these properties to dynamical systems which are not necessarily Hamiltonian.
Although the BHW Lagrangian flow \eqref{eq:hamiltonian} is a Hamiltonian system, we introduce the generalized framework since it emphasizes how the properties of integrable systems can be understood to arise from symmetries of physical systems, rather than from structures or geometry particular to Hamiltonian systems.

To connect dynamics with geometry, we will use the fact that vector fields on manifolds can be associated with first-order differential operators; for example, the convective derivative
\begin{equation*}
    D_t := \partial_t + u(x,y,t) \partial_x + v(x,y,t) \partial_y
\end{equation*}
is associated to the vector field \(\hat{\mathbf{t}} + u(x,y,t) \hat{\mathbf{x}} + v(x,y,t) \hat{\mathbf{y}}\), where \(\hat{\mathbf{t}}, \hat{\mathbf{x}}, \hat{\mathbf{y}}\) are unit vectors pointing in the time, \(x\), and \(y\) directions respectively.
Notice that here the manifold is the extended phase space \(\mathbb{R}^3\) which includes time as a coordinate, on which \(\mathbf{u}(x,y,t)\) is an autonomous vector field.
The Lagrangian trajectories \((X(t),Y(t),t)\) in the extended phase space are integral curves of \(D_t\).
In the following, we will not distinguish between vector fields and their first-order differential operators, and furthermore we will assume all functions are smooth.
All vector fields will also be assumed to be autonomous, which can be accomplished by extending the phase space if necessary.

Given two vector fields \(X,Y\), we say that they \textit{commute} if they commute as first-order differential operators, i.e. \([X,Y](f) := X(Y(f)) - Y(X(f)) = 0\) for any sufficiently differentiable scalar function \(f\).
It is known that the commutator of two vector fields gives another vector field \([X,Y]\), which satisfies \([X,Y]=0\) if the vector fields commute as operators.
Commutation is closely associated with the notion of symmetry.
For example, the velocity field \(\mathbf{u}=\mathbf{u}(x,y,t)\) being translationally invariant in the coordinate \(y\) is equivalent to the condition \([D_t,\partial_y] = 0\).
Noting that translation by \(r\) in the \(y\) direction can be formally associated with the operator exponential \(\exp(r\partial_y)\), we identify \(\partial_y\) as the \textit{infinitesmal generator} of the continuous symmetry corresponding to translation by \(y\).

In Hamiltonian systems, each conserved quantity gives rise to a continuous symmetry of the system corresponding to a commuting vector field.
To illustrate this, consider a time-autonomous canonical Hamiltonian system on a symplectic manifold of dimension \(2n\) with Poisson bracket \(\{\cdot, \cdot\}\).
Such systems are Liouville integrable when there exist \(n\) independent integrals of motion \(I_1 = H, I_2, ..., I_n\) satisfying \(\{I_i, I_j\} = 0\) for each \(i,j\).
These integrals of motion gives rise to \(n\) Hamiltonian vector fields \(X_1,..., X_n\) defined by \(X_j(f) := \{f, I_j\}\) for any phase space function \(f\).
The condition \(\{I_i,I_j\} = 0\) is then equivalent to a commutation condition \([X_i,X_j] = 0\), suggesting that the vector fields can be interpreted as generators for an \(n\)-dimensional Abelian Lie group (i.e. commuting continuous symmetry group).
Furthermore, the Poisson bracket condition gives \(X_j(I_i) = 0\) for all pairs \(i,j\), meaning the \(I_i\) are common integrals of motion for the vector fields \(X_j\).

Given these integrals of motion, the Liouville-Arnold theorem states that in the neighborhood of compact (i.e. closed and bounded) common level sets of the integrals of motion \(I_1,..., I_n\), there exists a foliation of the phase space into invariant tori.
Note the independence condition is equivalent to an assertion that most of the common level sets are \textit{regular}, i.e. the vectors \(\nabla I_1, ..., \nabla I_n\) are linearly independent everywhere on each level set.
Furthermore, the theorem states that there exist action-angle coordinates labeling these invariant tori, in which motion generated by the Hamiltonian \(H\) will be quasiperiodic.
The proof of the Liouville-Arnold theorem uses the fact that all compact \(n\)-dimensional Abelian Lie groups are isomorphic to \(\mathcal{U}(1)^{n}\), i.e. the set of rigid rotations on the \(n\)-torus \cite{Arnold1989}.
The Hamiltonian flow \(X_1\) is a linear combination of the generators of rigid rotation in each angle variable, leading to the quasiperiodic flow.

Through this lens of symmetry, an analogous \textit{Non-Hamiltonian version of Liouville's Theorem}, also known as the Generalized Liouville Theorem, can be proven for flows which are not necessarily Hamiltonian \cite{Bogoyavlenskij1998,Zung2016}.
Summarizing, the theorem says if an \(n\)-dimensional autonomous flow \(D_t\) has \(p\) commuting symmetries and \(r\) common integrals of motion to those symmetries with \(p+r=n\), then in the neighborhood of compact regular level sets of the integrals of motion, the phase space foliates into invariant tori.
Furthermore, there will exist \(p\) action and \(r\) angle coordinates (not necessarily canonically conjugate) labeling these invariant tori such that the Lagrangian trajectories of \(D_t\) will undergo quasiperiodic motion in the angle variables.
In physical terms, this means that any flow with a sufficient number of commuting symmetries and integrals of motion is necessarily laminar, even if the flow is not Hamiltonian.

Despite its name, the generalized theorem also applies in the Hamiltonian case, where the Poisson bracket greatly facilitates the search for these commuting symmetries and common integrals of motion.
The proof of the generalized theorem is exactly analogous to the proof of the Hamiltonian version, see the discussions in the given references.
A more precise discussion of this theorem as well as its requirements is left to Appendix \ref{app:integrability}.

\subsection{Laminar flow and fluid parcel relabeling symmetries}

Motivated by the fact that the PV is an approximate integral of motion in the BHW equation, we state the following proposition, which rephrases the previously reported results of \cite{Mezic1994,Haller1998} using the framework of Generalized Liouville integrability:
\begin{prop}[Lagrangian flow integrability in 2d hydrodynamic systems] \label{prop:2d_integrability_text}
    Let \((\mathbf{u},h,q)\) be a tuple consisting of a velocity field \(\mathbf{u}(x,y,t)\), a positive scalar function \(h(x,y,t) > 0\), and a scalar function \(q(x,y,t)\).
    
    Suppose \((\mathbf{u},h,q)\) satisfies a mass continuity equation and a scalar conservation equation for \(q\),
    \begin{subequations} \label{eq:2d_integrability}
        \begin{gather} 
            D_t h + h (\nabla \cdot \mathbf{u}) = 0, \\
            D_t q = 0.
        \end{gather}
    \end{subequations}
    
    Then the vector field (i.e. first-order differential operator) \(L_q\) defined by
    \begin{equation} \label{eq:lq}
        L_q f := \frac{(\partial_y q, -\partial_x q)}{h} \cdot \nabla f
    \end{equation}
    satisfies \([D_t,L_q] = 0\). In particular, if \(\nabla q \neq 0\) except possibly on a set of measure zero, then \((D_t, L_q, q)\) forms a system of 2 commuting symmetries and a common integral of motion sufficient for the Generalized Liouville theorem to apply.
\end{prop}
Note that the theorem here is stated for a possibly compressible flow \(\mathbf{u}\).
In the case of an incompressible flow, one can take \(h(x,y,t) = 1\).
Additionally, note that \(q\) in the proposition is an arbitrary singly-valued smooth function that serves as the integral of motion.
Some care needs to be taken when using the PV \(q+\kappa x\) as the integral of motion, as it is not singly-valued on the doubly-periodic domain.
This proposition has as a near-immediate corollary:

\begin{thm}[Lagrangian flow integrability of time-periodic solutions to BHW] \label{thm:chm_integrability}
Let \((q(x,y,t), \allowbreak n(x,y,t))\) be a smooth time-periodic solution to the BHW equations \eqref{eq:plasma_model} with \(\mu=0\) on a doubly-periodic domain. If \(\nabla(q+\kappa x) \neq 0\) except possibly on a set of measure zero, the Lagrangian flow of this solution defined by \eqref{eq:hamiltonian} will be Liouville integrable.
\end{thm}

In physical terms, Proposition \ref{prop:2d_integrability_text} along with Theorem \ref{thm:chm_integrability} imply that the Lagrangian flow of smooth time-periodic solutions to the inviscid BHW equations is necessarily laminar due to the existence of the conserved PV.
An analogous proposition for 3d time-dependent flows that generalize the results of \cite{Mezic1994}, as well proofs for the propositions and the theorem, are left to Appendix \ref{app:integrability}.
Examples of hydrodynamic systems with non-Hamiltonian Lagrangian flows satisfying similar integrability properties are given in Appendix \ref{app:integrability} as well.

Here, we remark on two key points in the theorem.
The first concerns the existence of the conserved scalar material invariant, the potential vorticity.
A wide variety of fluid and plasma systems have been shown to have a PV or PV-like invariant \cite{Muller1995,Gurcan2015}.
The existence of such conserved quantities can be seen as a consequence of Noether's theorem applied to the \textit{fluid parcel relabeling symmetry} of the infinite-dimensional Lagrangian field theory formulation of various fluid and plasma systems \cite{Salmon1988,Padhye1996}.
Such symmetries are also linked to Casimirs of functional Poisson brackets in the infinite-dimensional Hamiltonian field theory formulation \cite{Morrison1982,Holm1985,Morrison1998}.
Note that the integrability considered here is for the finite-dimensional system of ordinary differential equations \eqref{eq:hamiltonian} which determine the Lagrangian trajectories, whereas the symmetry originates from a property of the infinite-dimensional PDEs that govern the fluid system.

Broadly speaking, the fluid parcel relabeling symmetry arises from the need to track the motions and properties of individual fluid parcels over time in the Lagrangian description of continuum systems.
To achieve this tracking, each parcel is initially assigned a label, typically its initial position, which then follows the parcel as it moves through the fluid.
Since these labels are arbitrarily assigned, changing the labels should in principle lead to equivalent equations for the motion of the fluid.
For a general continuum system described by a Lagrangian field theory, the particle relabeling symmetry refers to the group of transformations on the particle labels that do not change the `physical' variables, such as the density or velocity, upon which the action depends.
Through Noether's theorem, this can often be shown to lead to the conservation of a PV or PV-like quantity.

Proposition \ref{prop:2d_integrability_text} gives a characterization for how this symmetry of infinite-dimensional fluid systems affects the finite-dimensional system of ODEs that define the Lagrangian flow.
The vector field \(L_q\) defines a flow corresponding to a continuous relabeling of fluid parcel labels.
In BHW, this flow will be perpendicular to \(\nabla (q+\kappa x)\), so parcels following this flow are confined to level sets of the PV.
The commutation condition then asserts that the operator ``rearrange positions of identical fluid parcels'' \(\exp(rL_q)\) commutes with the operator ``evolve along Lagrangian trajectories'' \(\exp(s D_t)\), giving a precise interpretation of the action of the fluid parcel relabeling symmetry on the Lagrangian trajectories.
Note that the statement of the proposition does not depend specifically on the fluid parcel relabeling symmetry, and if the integral of motion arises from a different symmetry, then the interpretation of \(L_q\) will also change.

The second key point concerns the physical interpretation of the requirements of non-degeneracy \(\nabla (q + \kappa x) \neq 0\) and time-periodicity on the solution, the latter of which is an especially restrictive condition.
To explain their necessity, note that in the extended phase space any sufficiently smooth flow can be endowed with `trivial' time-dependent integrals of motion: given an arbitrary smooth function \(a_0(\mathbf{x})\), the PDE
\begin{equation*}
    \partial_t a(\mathbf{x},t)  + \mathbf{u} \cdot \nabla a(\mathbf{x},t) = 0; \qquad a(\mathbf{x},t_0) = a_0(\mathbf{x})
\end{equation*}
has a solution \(a(\mathbf{x},t)\) which is formally an integral of motion satisfying definition \ref{def:integrability}.
This holds even if \(\mathbf{u}\) generates a chaotic flow.

Since integrals of motion are constant along trajectories, the trajectories are necessarily confined to individual level sets of the integrals of motion.
If \(a(\mathbf{x},t)\) is a constant function, then \(a\) has only one level set consisting of the entire domain, so the existence of this trivial integral of motion does not provide any information or constraints on the motion of the fluid parcels.
The non-degeneracy requirement \(\nabla (q + \kappa x) \neq 0\) ensures that the integrals of motion are not locally constant and thus have non-trivial `laminar' level sets that confine the trajectories.

Then, given a non-degenerate integral of motion, the time-periodicity requirement is necessary to prevent its level sets from `wandering'.
Namely, in the time-dependent case the level sets of \(a(\mathbf{x},t)\) could develop small-scale meanders that become arbitrarily complicated over time, again failing to provide a meaningful constraint on the motion of the fluid parcels.
Time-periodicity ensures that the level sets of the integral of motion \(q+\kappa x\) do not become arbitrarily complex in time, and periodically return to their original shape.
Additionally, the time-periodicity ensures that the level sets are bounded in the time direction and hence compact, which is necessary for Liouville integraility to guarantee quasiperiodic motion confined to invariant tori.

\subsection{Near-integrability and weak chaos in eigenmode flows}

The requirement in Theorem \ref{thm:chm_integrability} for exact time-periodic solutions is quite restrictive.
Here, we will show how relaxing the requirement to time-periodic approximate solutions results in a formal near-integrability condition, applicable to drift-wave eigenmodes in Hasegawa-Wakatani.
Given that integrable flows are laminar flows, this implies that zonal flows plus eigenmodes should generally be expected to induce ``nearly-laminar'' Lagrangian flows.

To begin, note that time-independent zonally-symmetric states with \(q=\bar{q}(x)\) and zonal flow profile \(U(x) = \bar{\varphi}^{\prime}(x)\) satisfy the requirements of Theorem \ref{thm:chm_integrability}, and therefore have integrable Lagrangian flows.
Now, consider perturbations to this background state given by
\begin{gather*}
    \varphi_{\epsilon}(x,y,t) = \bar{\varphi}(x) + \epsilon \tilde{\varphi}(x,y,t), \\
    q_{\epsilon}(x,y,t) = \bar{q}(x) + \epsilon \tilde{q}(x,y,t)
\end{gather*}
where \(\epsilon\) is a formal perturbation parameter, and where \(\tilde{\varphi}\) and \(\tilde{q}\) are arbitrary functions at this stage.

Generically, \(q_{\epsilon}\) will no longer correspond to an integral of motion of the time-dependent flow induced by the perturbed potential \(\varphi_{\epsilon}\), and instead satisfies the condition
\begin{equation} \label{eq:near_integrability}
    D_{t,\epsilon} (q_{\epsilon} + \kappa x) := \partial_t q_{\epsilon} + \{\varphi_{\epsilon}, q_{\epsilon} + \kappa x\} = \epsilon R_1 + \epsilon^2 R_2
\end{equation}
where
\begin{equation} \label{eq:near_integrability_r1}
    R_1 := \partial_t \tilde{q} + U(x) \partial_y \tilde{q} - (\bar{q}'(x) + \kappa) \partial_y \tilde{\varphi}.
\end{equation}

Suppose we wish to find time-periodic perturbations which are near-integrable in the sense of setting the remainder terms on the right-hand side of \eqref{eq:near_integrability} close to \(0\).
One way to do this is to set \(R_1 = O(\mu)\), where \(\mu\) will typically be very small.
Comparing this with \eqref{eq:eig}, we see that the Fourier transform in \(t\) and \(y\) of this near-integrability condition exactly corresponds to one of the equations which determine the eigenfunctions \eqref{eq:eig_q}.
Thus, we conclude that the flow induced by a superposition of zonal flows plus drift-wave eigenmodes has a \textit{formal near-integrability property}, with an approximate integral of motion \(q_\epsilon + \kappa x\).
This is analogous to the near-integrability property derived in \cite{cao2023rossby} for Rossby waves.

Note there is an implicit assumption in this property that the eigenmodes are non-singular, as any perturbative assumption would break down if the amplitude of the eigenmode envelope diverges at any point.
Furthermore, this formal property on its own is not sufficient to guarantee that the flow is actually nearly-integrable.
For example, significant overlap between the resonances associated with the eigenmodes could lead to Lagrangian flow which is primarily chaotic.
However, as we will see in the following sections, the turbulent flow observed in the DNS does appear to be weakly chaotic, suggesting that some form of near-integrabiilty holds in the weakly turbulent regime studied here.

Recalling the discussion on the role of time-periodicity in Theorem \ref{thm:chm_integrability}, the requirement for time-periodicity here is necessary for concepts from integrability such as invariant tori to meaningfully constrain the flow.
In particular, this suggests the importance of neutrally stable eigenmodes in organizing the Lagrangian flow, as these modes have a purely time-periodic evolution.
Since perturbations only need to approximately satisfy the eigenmode equations, marginally unstable modes with amplitude taken to be fixed are also eligible for this near-integrability property.
Modes with such stability are expected to be generic in weakly driven turbulent systems due to the tendency for background gradients to relax towards marginality, in line with discussions in earlier works such as \cite{DelCastilloNegrete1993}.

In practice, there are several issues applying the formal near-integrability property to the drift waves observed in the DNS.
First, given the discussion in section \ref{sec:eigenmodes} it is unlikely that the formal perturbation parameter \(\epsilon\) can be considered small.
More subtly, the dominant perturbations to the eigenfunction flow tend to be non-periodic.
The presence of the length scale \(\rho_s = 1\) determined by the gyroradius in the Hasegawa-Wakatani equations tends to organize the flows locally, as opposed to globally.
For example, the drift-waves present in the \(-y\) direction zonal bands tend to organize into wavepackets which experience significant distortions to their shape and amplitude before completing one poloidal transit (see Fig.~\ref{fig:Zonal-transport}).
This makes it difficult to phrase near-integrability in terms of the survival of invariant tori, as is typical in applications of KAM theory.
To address these issues, in the next section we turn instead to the characterization of the transition from integrability to chaos in the Lagrangian flows.

\subsection{Separatrix splitting and and chaotic tangles}
It has been well established, especially in the context of Hamiltonian flows in the plane, that perturbed integrable systems can proceed to chaos through the emergence of stochastic regions near separatrices in the unperturbed flow.
In the weakly chaotic regime, which typically occurs before there is significant island overlap, this process can be characterized by the phenomenon known as \textit{separatrix splitting}, see e.g. \cite{Balasuriya2016} for a comprehensive text relevant to fluid flows.
Here, we first summarize the key points in the theory for planar flows and discuss some issues which arise in aperiodic finite-time settings relevant to the DNS snapshots.
However, the theory is more general than the planar case, possibly applying to \(n\)-dimensional dynamical systems.

To illustrate the concepts, recall that an autonomous planar Hamiltonian flow is always integrable, and hence laminar.
Regions of the plane with different flow topology, which will have different action-angle coordinates, are separated by curves known as separatrices.
These separatrices consist of the \emph{stable and unstable manifolds} (SUMs) associated with hyperbolic (saddle-like) fixed points, also referred to as X-points.
Here, the stable and unstable manifolds are defined as the set of Lagrangian flow trajectories $\mathbf{X}(t)$ which exponentially converge towards their associated hyperbolic fixed point as \(t \to \infty\) or \(t \to -\infty\) respectively.
The basic idea behind separatrix splitting is that in the integrable case, the SUMs coincide and form smooth, unbroken separatrix curves.
In the presence of a weak perturbation the SUMs will persist, but in general will no longer coincide and instead will form a fractal \textit{chaotic tangle}.
These tangles no longer cleanly separate regions with differing flow topology, leading to the formation of the stochastic regions near the unperturbed separatrices.

A key concept in this picture of separatrix splitting is the notion of \textit{hyperbolic trajectories} and their associated SUMs.
Hyperbolic trajectories correspond to perturbed versions of the stationary trajectories associated with the X-points.
Here, hyperbolicity refers to the behavior of the linearized Lagrangian flow around the trajectory. 
To illustrate this, consider a trajectory \(\mathbf{X}(t)\) of the characteristic equations \eqref{eq:hamiltonian}.
The dynamics of an infinitesimally perturbed trajectory \(\mathbf{X}(t) + \boldsymbol{\xi}(t)\) will evolve through the linearized equations for \(\boldsymbol{\xi}(t)\),
\begin{equation}\label{eq:lin_coeff}
    \dot{\boldsymbol{\xi}}(t) = A(t) \boldsymbol{\xi}(t) \coloneqq \left[\frac{\partial \mathbf{u}}{\partial \mathbf{x}}(\mathbf{X}(t),t)\right] \boldsymbol{\xi}(t).
\end{equation}
Here \(A\) is the strain-rate tensor evaluated along the trajectory with components \(A_{ij} = \frac{\partial u_i}{\partial x_j}\).

The key idea is that if \(A\) has contracting  (resp. expanding) directions, trajectories perturbed by \(\boldsymbol{\xi}(t)\) in these directions will remain close to the original trajectory \(\mathbf{X}(t)\) forward (resp. backward) in time, mimicking key features of hyperbolic fixed points.
Note that at each instant in time, the Okubo=Weiss criterion \cite{Okubo1970,Weiss1991} \(Q := \varphi_{xy}^2 -\varphi_{xx} \varphi_{yy} = -\det A\) determines if the eigenvalues of \(A\) are real (\(Q>0\), hyperbolic) or imaginary (\(Q<0\), elliptic), and hence whether the flow is expanding/contracting at that instant in time.
However, \(A(t)\) may change as the trajectory samples different regions of the flow, so knowing just the character of the eigenvalues of \(A(t)\) at each instant in time may not necessarily be enough to determine if nearby trajectories converge or diverge over longer intervals of time.
We cover a more precise definition of hyperbolic trajectories in Section~\ref{subse:num-sums}.

The existence of contracting and expanding directions allows for a definition of SUMs associated with these hyperbolic trajectories.
The stable (resp. unstable) manifold of a hyperbolic trajectory consists of the set of trajectories which exponentially converge towards the hyperbolic trajectory as \(t \to \infty\) (resp. \(t \to -\infty\)).
The SUMs associated with these hyperbolic trajectories can be understood as perturbed version of the separatrices associated with the X-points.
In the 3-dimensional extended phase space that includes time as a coordinate, the SUMs take the form of 2-dimensional surfaces which meet at 1-dimensional curve corresponding to the hyperbolic trajectory.
In time-periodic settings, it can be rigorously shown using Melnikov theory that if the stable and unstable manifolds have at least one additional transversal intersection, they will form a fractal chaotic tangle \cite{Wiggins2003}.
Many other properties expected of chaotic systems, such as the generation of topological entropy, can also be rigorously proven in this case.

One application of chaotic tangles familiar to magnetically confined fusion plasmas is the description of electron transport due to stochastic magnetic fields that form near the X-point of the separatrix corresponding to the last closed flux surface \cite{Evans2004}.
The basic idea is that the free-streaming trajectories of electrons along magnetic field lines correspond to integral curves of the magnetic field \(\mathbf{B}(R,Z,\theta)\), here expressed in cylindrical coordinates aligned with the central axis of the torus.
For time-periodic flows \(\mathbf{u}(x,y,t)\), a close geometric analogy can be drawn by identifying \(\hat{\mathbf{t}} + \mathbf{u}\) with \(\mathbf{B}\), and the periodic coordinate \(t\) with \(\theta\).

Moving beyond time-periodic flows, there have been successful extensions of this picture of SUMs to aperiodic, finite-time flows, which are relevant to the locally organized vortical flows observed in the DNS.
For example, there are extensions of the perturbative Melnikov theory to finite-time aperiodic settings \cite{malhotra1998geometric,Sandstede2000}, which provide analytical formulas for the perturbed position of the SUMs.
Another approach, rather than identifying individual SUMs, instead looks at finite-time Lyapunov exponents (FTLEs).
As FTLEs are spatially-varying fields, they can provide more detailed information about the the hyperbolic behavior of nearby trajectories \cite{Finn2001,Haller2002,Haller2015}.
Note that ridges in the FTLE can also be used to construct approximate SUMs, see the review in \cite{Haller2015}.
In this work, we will proceed with the finite-time theory of distinguished hyperbolic trajectories (DHTs) defined in \cite{ide2002distinguished}.
This has the convenience of being framed in terms of individual hyperbolic trajectories and material curves which approximate the SUMs to these trajectories.

A key property of the SUMs that links them to transport is the fact that they are material curves, i.e. curves which move with the Lagrangian flow.
Since Lagrangian trajectories are uniquely determined by their position at any point in time (assuming a sufficiently smooth flow), fluid parcels cannot ever cross the SUMs.
Thus, the only transport which can occur across SUMs is diffusive.
This means the SUMs can act as transport barriers, particularly for the PV which is a scalar material invariant in the absence of viscosity \(\mu\).
Broadly speaking, SUMS viewed in this way fall under the category of Lagrangian coherent structures (LCSs), such as the hyperbolic LCSs in \cite{Haller2015}.

Lastly, When the SUMs have multiple transversal intersections, as is the case when chaotic tangles form, the SUMs enclose regions of fluid between these intersections known as `lobes'.
Through a process known as the `turnstile', these lobes mediate the exchange of fluid parcels across the former separatrix, leading to turbulent transport.
We delay a detailed discussion of the lobe dynamics to section \ref{sec:lobedynamics}, although the reader is referred to \cite{meiss2015thirty} for a comprehensive review of turnstile and lobe dynamics.
In the next section, we will demonstrate how this route to chaos involving chaotic tangles is then linked to intermittency in transport and dissipation observed in the Dimits shift regime.

\section{Persistence of nearly-integrable flow topology in weakly turbulent regimes} \label{sec:sums}

In this section, we study the persistence of coherent large-scale
structures by computing their hyperbolic trajectories and associated
stable and unstable manifolds. In order to do this, we focus on the
Lagrangian transport equations \eqref{eq:hamiltonian} of the fluid system so that the typical
vortical structures can be characterized by the stable and unstable
manifolds developed from the hyperbolic Lagrangian trajectories. In
addition, we compare the results from both the full flow solution
and the reconstructed eigenfunctions \eqref{eq:eig}. This shows how stable/unstable manifold theory provides an
understanding of the mechanisms which mediate transport
and organize the observed large-scale structures in the Dimits shift regime of BHW turbulence.

\subsection{Numerical strategies to compute hyperbolic trajectories and the associated stable and unstable manifolds}\label{subse:num-sums}

In the first place, we describe the general strategy adopted to locate
the hyperbolic trajectories from the advection flow field data, then
compute the stable and unstable manifolds based on the identified
hyperbolic trajectories. The hyperbolic trajectories and manifolds
will be computed from the BHW model data for the analysis of anomalous
transport of Lagrangian states and the particle flux transition in
Dimits shift.
The velocity fields along $x$ and $y$ directions \(\mathbf{u}=(u,v)\) for the computation will come from either the true flow field from direct numerical simulation of \eqref{eq:plasma_model} or the reconstructed field \eqref{eq:eig} from the
eigenfunctions.
A constant velocity $V_c$ is introduced tracking the moving vortices by the flow field along $y$ direction. 
We consider a fixed time window $t\in\left[0,T\right]$.

We first summarize the unified approach for computing the hyperbolic
trajectory and corresponding manifolds in the following three steps:
\begin{itemize}
\item \emph{Step I}: locate the \emph{instantaneous stagnation points} (ISP)
at each time instant $t$ based on a constant moving coordinate velocity  $V_c$
along $y$ direction;
\item \emph{Step II}: find the \emph{distinguished hyperbolic trajectory}
(DHT) based on the iterative procedure from the initial guess by connecting
the ISPs;
\item \emph{Step III}: compute the \emph{stable and unstable manifolds}
(SUMs) from the the recovered DHT by integrating the Lagrangian equation
forward and backward in time.
\end{itemize}
First, the ISPs $\left(X_{c}\left(t\right),Y_{c}\left(t\right)\right)$
at time instant $t$ are defined by the critical saddle points in
the Eulerian flow field such that
\begin{equation}
u\left(X_{c},Y_{c},t\right)=0,\;v\left(X_{c},Y_{c},t\right)=0,\quad\mathrm{and}\quad\frac{\partial\left(u,v\right)}{\partial\left(x,y\right)}=\det\left(\begin{bmatrix}u_{x} & u_{y}\\
v_{x} & v_{y}
\end{bmatrix}\right)<0.\label{eq:isp}
\end{equation}
The hyperbolicity is determined by one stretching direction with positive
Lyapunov exponent and one compressing direction with negative exponent
in the above determinant. More precisely, we can define the \emph{hyperbolic trajectory} of the time-dependent flow following \cite{malhotra1998geometric} as:
\begin{defn}
A trajectory $\boldsymbol{\gamma}(t)$ of the Lagrangian flow \eqref{eq:hamiltonian} is called a hyperbolic trajectory if the associated linearized system \eqref{eq:lin_coeff}
\begin{equation}\label{eq:dich}
    \dot{\boldsymbol{\xi}}(t) =  \left[\frac{\partial \mathbf{u}}{\partial \mathbf{x}}(\boldsymbol{\gamma}(t),t)\right] \boldsymbol{\xi}
\end{equation}
has an exponential dichotomy. That is, there exists a projection $P$ and constant $K,\lambda>0$ such that the fundamental matrix $Y(t)$ of \eqref{eq:dich} satisfies
\begin{equation}\label{eq:dich1}
\begin{aligned}\left\Vert Y\left(t\right)PY^{-1}\left(t\right)\right\Vert  & \leq K\exp\left(-\lambda\left(t-s\right)\right),\quad t\geq s,\\
\left\Vert Y\left(t\right)\left(Id-P\right)Y^{-1}\left(t\right)\right\Vert  & \leq K\exp\left(\lambda\left(t-s\right)\right),\quad t<s.
\end{aligned}
\end{equation}
\end{defn}
In the BHW model, however, one key feature is the
persistent zonal flow structure along $y$ direction. As the result,
the ISPs defined above no longer remain in a small neighborhood but
move along the $y$ direction with a varying zonal velocity changing
along the $x$-axis (such as the example in Fig.~\ref{fig:dns_snapshot}). This
impacts the identification of proper locations of the hyperbolic trajectories.
Usually, the hyperbolic trajectories are required to remain in a bounded
region during the entire time. In order to avoid this subtlety with
the traveling ISPs, we introduce a new coordinate system by the phase
velocity from the eigenfunctions. In this way, we are able to focus
on each localized coherent vortices around a specific region of the fluid
field.

\begin{figure}
\includegraphics[scale=1.2]{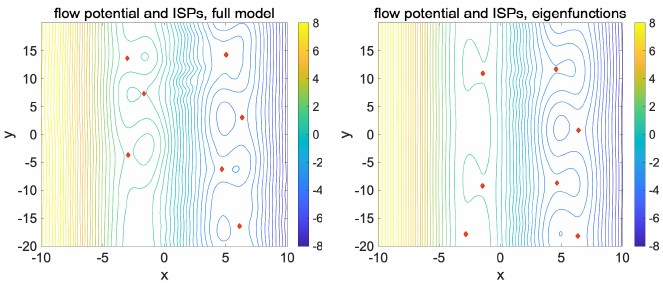}

\caption{The flow potential functions and the corresponding ISPs from the shifted
states under the new moving coordinates in (\ref{eq:shifted_fields}).
Results using both the full model data (left) and the reconstructed
field from the eigenfunctions (right) are compared.\label{fig:The-flow-potential}}

\end{figure}

In particular, we adopt the following precise definition \cite{ide2002distinguished}
of the \emph{distinguished hyperbolic trajectory} for a weakly turbulent
flow field:
\begin{defn}
Assume there exists an invertible change of coordinate $\left(x^{\prime},y^{\prime}\right)$
from the original coordinate $\left(x,y\right)$. If the hyperbolic
trajectory $\boldsymbol{\gamma}^{\prime}$ under the new coordinate
remains in a neighborhood $\mathcal{B}$ for the entire time, and all other trajectories starting in $\mathcal{B}$ leave $\mathcal{B}$ in finite time, the
trajectory $\boldsymbol{\gamma}$ under the original coordinate is said
to be a distinguished hyperbolic trajectory of (\ref{eq:hamiltonian}).
\end{defn}
According to the above definition of DHT, we introduce the new moving
coordinate with a constant velocity $V_{c}$ focusing on a specific
localized region of the fluid field with varying zonal velocity, that
is,
\begin{equation}
x^{\prime}=x,\quad y^{\prime}=y-V_{c}t,\quad t^{\prime}=t.\label{eq:moving_coord}
\end{equation}
In this way, we find the time-invariant states under the new moving
coordinate system such that certain DHT in the above definition can
be discovered in the proper region with a characteristic moving speed
$V_{c}$ along the $y$-direction
\begin{equation}\label{eq:shifted_fields}
\begin{aligned}
\varphi^{\prime}\left(x,y^{\prime},t\right)=&\varphi\left(x,y^{\prime}+V_{c}t,t\right)-V_{c}x,\\ u^{\prime}\left(x,y^{\prime},t\right)=&u\left(x,y^{\prime}+V_{c}t,t\right),\\
v^{\prime}\left(x,y^{\prime},t\right)=&v\left(x,y^{\prime}+V_{c}t,t\right)-V_{c}.
\end{aligned}
\end{equation}
The key issue becomes to determine the proper constant moving velocity
of the $y$ coordinate $V_{c}$ so that the local boundedness of the
hyperbolic trajectory is guaranteed. Different values of $V_{c}$ lead to
different ISPs due to the shifting locations of zero points
for $v^{\prime}$.
Motivated by the strong linear character of the coherent vortices, here we pick \(V_c\) to be the phase velocity \(u_{ph}\) of the eigenmodes.
In Fig.~\ref{fig:The-flow-potential},
we plot the potential function $\varphi^{\prime}$ in (\ref{eq:shifted_fields})
under the characteristic velocity $V_{c}=0.8$ which represent the
focused region around $x\in\left[0,10\right]$ depicted in Fig.~\ref{fig:Zonal-transport}.
The ISPs are identified by locating the critical points with $u^{\prime}=0$
and $v^{\prime}=0$. In addition, we observe that the reconstructed
field from eigenfunctions also reveal the locations of the ISPs in the focused region $x\in\left[4,8\right]$. In another region around $x\in\left[-5,0\right]$ where the zonal flow possesses a different characteristic velocity, the ISPs do not agree with the definition of remaining in a small neighborhood during the evolution thus cannot be used to discover the proper DHTs.

Next, we provide more explanations on the detailed numerical strategies
to compute the DHTs then the associated SUMs using the identified
ISPs above as initialization states.

\subsubsection{Iterative method for identifying the distinguished hyperbolic trajectories}

With the ISPs as the critical points of \eqref{eq:shifted_fields} identified, a practical numerical
method is needed to recover the DHTs defined from the connected ISPs
from the previous step. In particular, the ISPs do not give the 
Lagrangian trajectories of the fluid velocity field, and also
could be very ill-posed with singular coefficient matrix in \eqref{eq:dich} and miss the accurate location of the hyperbolic
trajectory. Thus an iterative algorithm is proposed to achieve the
converging DHT starting from a careful choice of the ISP trajectory
$\mathbf{X}^{0}\left(t\right)$ as the initialization and updating
over a proper time window $\left[0,T\right]$ forward in time along the
stable direction and backward in time along the unstable direction. 

In particular, at the $i$-th iteration step, assume we have the trajectory
$\mathbf{X}^{i}\left(t\right)$. We introduce the new disturbance
variable $\boldsymbol{\xi}=\mathbf{X}-\mathbf{X}^{i}$, where $\mathbf{X}$
is the target distinguished hyperbolic trajectory. Then we can derive
the explicit dynamical equations for $\boldsymbol{\xi}$
\begin{equation}
\dot{\mathbf{\boldsymbol{\xi}}}=\dot{\mathbf{X}}-\mathbf{\dot{X}}^{i}=A^{i}\left(t\right)\boldsymbol{\xi}+\mathbf{f}^{i}\left(t\right),\label{eq:iter_dht}
\end{equation}
with the linearized coefficients and additional forcing in the following
explicit forms
\[
A^{i}\left(t\right)=\frac{\partial\mathbf{u}}{\partial\mathbf{x}}\left(\mathbf{X}^{i}\left(t\right),t\right),\quad\mathbf{f}^{i}\left(t\right)=\mathbf{u}\left(\mathbf{X}^{i}\left(t\right),t\right)-\mathbf{\dot{X}}^{i}\left(t\right)+\mathrm{h.o.t.}
\]
In practical implementations, the time differentiation $\mathbf{\dot{X}}^{i}$
is computed by a finite time difference scheme (the central
difference scheme is used in our applications). The higher-order terms in $\mathbf{f}^{i}$
is then neglected. In order to make $\mathbf{X}$ converge to a proper
DHT, the linearized matrix $A$ is required to contain one positive
and one negative eigenvalue. This enables the computation of the hyperbolic
trajectory be integrating the equations forward and backward in
time for the stable and unstable branch correspondingly as the `exponential
dichotomy' \eqref{eq:dich1}. The convergence of the DHTs is shown in \cite{ide2002distinguished}.
On the other hand, if it is found that the coefficient matrix $A^{i}$
goes to near singular values to break the hyperbolicity, it implies
the break down of the corresponding hyperbolic trajectory in time.

\begin{figure*}
\includegraphics[width=\linewidth]{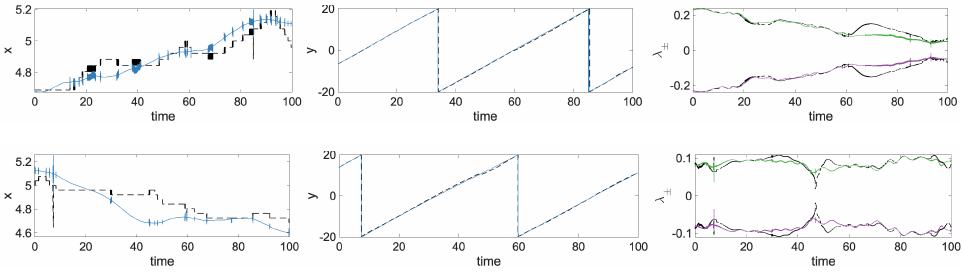}

\caption{Converged DHTs in $x$ (left) and $y$ (middle) components starting from different ISPs of the BHW model as well as the two eigenvalues of the linearized coefficient (right). Black
dashed lines are the initial states and the solid lines are the final
converged state.\label{fig:Converged-dht}}
\end{figure*}
Fig.~\ref{fig:Converged-dht} plots the converged DHTs through the
iterations from the different ISP trajectories in a time window $\left[0,100\right]$.
The first two columns show the time trajectories of the converged
DHTs compared with the initial states, while the third column shows
the two eigenvalues $\lambda_{\pm}$ of the linearized coefficient
$A\left(t\right)$. It can be seen that the $y$ component of the
DHTs are moving at a constant speed consistent with our assumption.
If the two eigenvalues of the linearized coefficient are away from
zero, the solution will converge to a nice final DHT. On the other
hand, if the linearized coefficient matrix becomes singular (with zero real part of eigenvalues) at
some time instant, hyperbolicity of the trajectory is violated then the resulting hyperbolic trajectory breaks to a
diverging trajectory. This infers that
the hyperbolic trajectory ceases to exist when there is stronger mixing, breaking the nearly-integrable character of the flow field.

\subsubsection{Computing stable and unstable manifolds according to a hyperbolic trajectory}

In the final step, we compute the SUMs according to the DHTs from the previous step. The strategy to achieve this is rather simple: we start at the initial (resp. final) point of the saddle point at the hyperbolic trajectory and extend it to a small segment along the stable or unstable eigendirection of the linearized coefficient matrix. Then we evolve this segment forward (resp. backward) in time with the flow velocity field. In particular, the unstable manifold starts from the initial position of the DHT solution at $t=0$ and is integrated forward in time, while the stable manifold starts from the final position of the same DHT at $t=T$ and is integrated backward in time. We use the 4-th order Runge-Kutta scheme for accurate time integration. The resulting surface gives the SUMs for the corresponding hyperbolic trajectory. 

In the practical algorithm, the initial segments of the stable and
unstable manifolds are a short line with only a few consecutive pointing
along the characteristic stable and unstable directions, that is,
the unstable eigenvector at the starting
time and the stable eigenvector at
the final time. The initial segments are then evolved in time by the
near turbulent velocity field according to the Lagrangian dynamics
\eqref{eq:hamiltonian}. The initial geometry of the segment will be stretched or compressed in time along the manifold. Thus the gaps between the adjacent points
will be gradually enlarged due to the advection flow field. A technical trick can be adopted here by effectively adding approximation points to the resolved manifolds \cite{ide2002distinguished}. Therefore,
at each time updating step, a gap inspection is first performed checking
the distance between the adjacent points on the SUMs. If an unacceptably
large gap is detected, a third-order interpolation is applied to fill
in the gaps between the two points. Similarly, if the adjacent points
are compressed too close to each other, the points will be redistributed
to make them evenly spaced on the manifolds.

\subsection{Lobe dynamics and intermittency in BHW} \label{sec:lobedynamics}

Here, we discuss the observations in the DHTs and the corresponding SUMs computed from the numerical strategy for the BHW model.
In particular, we show that intermittency in the turbulent dissipation and transport can be linked to the dynamics of the lobes linked to the SUMs originating from a splitting of the coherent vortex separatrices.
The dynamical evolution of the lobes provides a way to understand the anomalous transport of PV across the nearly-laminar zonal flows and vortex structures in the Dimits regime observed in the BHW model.

\begin{figure*}
	\centering
    \includegraphics[scale=1.]{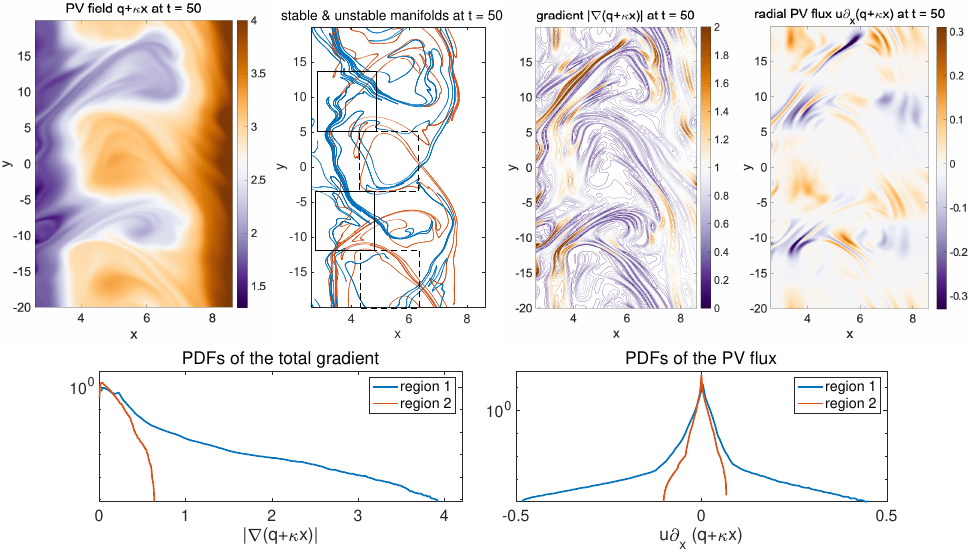}
    
    \caption{First row from left to right: i) profile of the PV $q+\kappa x$; ii) intersecting
    stable (blue) and unstable (red) manifolds from the full model simulated
    data; iii) the corresponding PV gradient field $\left|\nabla\left(q+\kappa x\right)\right|$; iv) radial PV flux divergence $u\partial_{x}(q+\kappa x)$. Second row: PDFs of the PV gradient field and radial PV flux divergence in two regions: region 1 denoted in solid blocks; and region 2 in dashed blocks  in the manifold plot.\label{fig:manifolds}}
\end{figure*}

In Fig.~\ref{fig:manifolds}, we show the computed SUMs according to the DHTs computed directly from the full model data.
In comparison, the associated fields for the PV $q+\kappa x$, the magnitude of PV gradient $\left|\nabla\left(q+\kappa x\right)\right|$, as well as the radial PV flux divergence $u\partial_x(q+\kappa x)$ are also plotted.
It can be observed that the intersecting SUMs accurately characterize the shapes and locations of the four major large-scale coherent vortices in the PV field.
Additionally, note that large values of the PV gradient and radial PV flux divergence coincide with the locations where the manifolds display strong folding structures, corresponding to regions with strong dissipation and mixing of the PV fluctuations.
To show a more quantitative characterization of the intermittency, we plot the PDFs of the total PV gradient $\left|\nabla\left(q+\kappa x\right)\right|$ and the radial PV flux divergence $u\partial_x(q+\kappa x)$ in the lower panel of Fig.~\ref{fig:manifolds}. The PDFs are plotted in two flow regions denoted by the solid and dashed blocked in the manifold plot. Region 1 focuses on the strongly mixing region where the SUMs intersect, while region 2 focuses on the large-scale coherent structures. Consistent with the previous qualitative discussion, fat tailed PDFs are observed in the strongly mixing region (region 1) where the lobe structures are  developed from the folding SUMs, while the coherent region (region 2) shows much shorter tails implying no intermittent transport among the coherent flow region.
In the following, we will briefly review some of the `turnstile' dynamics of the `lobes' associated with the chaotic tangles, and discuss how these features can be seen as a consequence.
See \cite{meiss2015thirty} for a more comprehensive review.

\begin{figure*}
	\centering
    \includegraphics[width=\linewidth]{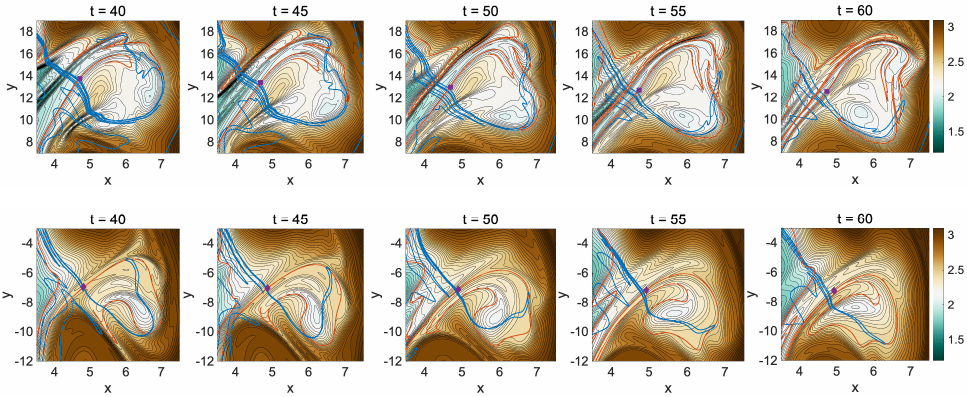}

    \caption{Tracking the time evolution of the homoclinic-like tangles of stable (blue)
    and unstable (red) manifolds of the same hyperbolic trajectory in the PV
    field $q+\kappa x$ of two localized coherent vortices. The hyperbolic trajectories are marked by purple
    dots.\label{fig:Tracking-manifolds}}
\end{figure*}

\begin{figure*}
	\centering
    \includegraphics[width=1.\linewidth]{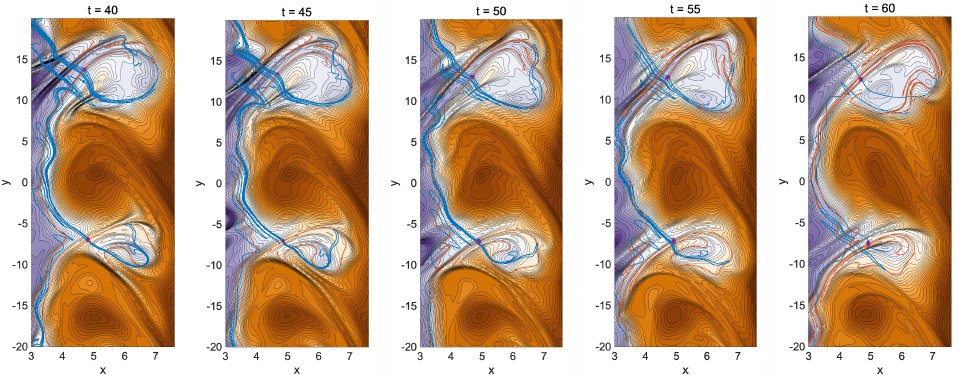}
    
    \caption{Tracking the time evolution of the heteroclinic-like tangles of stable
    (blue) and unstable (red) manifolds of two hyperbolic trajectories in the global PV field $q+\kappa x$. The hyperbolic trajectories are marked
    by purple dots.\label{fig:Tracking-manifolds-1}}
\end{figure*}

To illustrate how the SUMs organize the time evolution of the flow near the coherent vortices, we plot local snapshots of the PV along with SUMs from two of the coherent vortices in Fig.~\ref{fig:Tracking-manifolds}.
Under the moving coordinate system with constant velocity $V_{c}$ in $y$ direction, the large vortices are localized in space and maintain their shape in time so that the large-scale coherent structures remain around an approximately fixed position in the computational domain.
The two rows correspond to homoclinic-like tangles associated with the two DHTs located in the upper and lower half region of the computational domain.
We also plot a larger snapshot of the PV along with the heteroclinic-like tangles that connect the two large-scale coherent vortices in Fig.~\ref{fig:Tracking-manifolds-1}.
In both figures, in addition to the primary vortex area enclosed by the SUMs, many `lobes' formed by secondary transverse intersections between the SUMs are also seen.

It is interesting to observe that in the first row of Fig.~\ref{fig:Tracking-manifolds}, the folded stable manifolds (in blue lines) are pushed away from the unstable manifold (in red lines) near the DHT.
This is due to the stretching property along the direction of the unstable manifold.
Similarly in the second row, the unstable manifold is pulled toward the hyperbolic point along the stable manifold as time goes forward.
The feature with one attracting direction and one repelling direction originates from the hyperbolicity of the DHTs.
This evolution of SUMs is also observed in the global manifold structure depicted in Fig.~\ref{fig:Tracking-manifolds-1}.
It can be observed that the folded stable manifolds are pushed away along the line of unstable manifolds, while unstable manifolds fold and accumulate along the other direction of the stable manifolds.

Note that the lobes formed by the folded unstable manifolds align extremely well with the elongated filaments of PV observed earlier by the unstable manifolds of the frozen flow.
Due to the area-preserving property of the flow, the lobes must stretch and distort transverse to their direction of motion as they accumulate towards an unstable manifold.
Through this process, the long and thin filaments are built up by the attraction to the unstable manifold.
The resulting filaments create large gradients in the PV, leading to the regions of strong dissipation.
Furthermore, the filaments often make large radial excursions, leading to significant radial transport of the PV.

Conversely, the lobes formed by the folded stable manifolds tend to cut radially across contours of the PV.
As they are pulled along the unstable manifold they become less folded, transporting ``fresh'' PV towards the unstable manifolds.
Eventually the section of the lobe consisting of the unstable manifold stretches and begins to fold, creating a new folded portion of the unstable manifold.
Since the unstable manifold cannot create a new crossing with the stable manifold as it folds, the unstable manifold tends to folds in the direction opposite from the stable manifold.
This creates an alternating pattern of incoming and outgoing lobes that cross the former separatrices, known as the turnstile.
This is seen to be linked to the alternating sign of the radial PV flux divergence observed in Fig.~\ref{fig:manifolds}.

\begin{figure*}
\centering
\includegraphics[width=1.\linewidth]{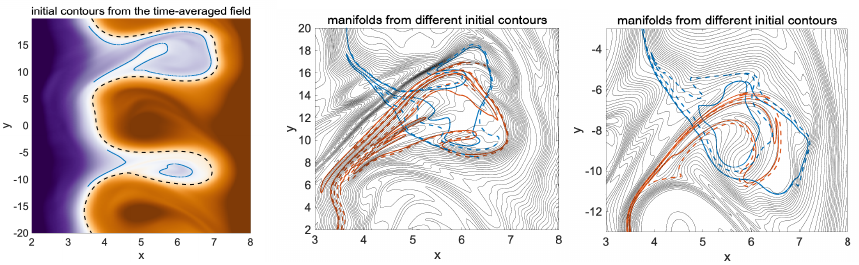}

\caption{Illustration of lobe invariance by computing the stable and unstable manifolds
from different initial contours. Different initial contours are represented
by solid and dashed lines and thee resulting SUMs are plotted in solid
and dashed lines accordingly.\label{fig:Illustration-of-shadowing}}
\end{figure*}

Furthermore, as an illustration of the robustness of the lobe structure under perturbations, we compute the manifolds developed from different initial contours in Fig.~\ref{fig:Illustration-of-shadowing}. 
Different from the initial segment in the previous results, we start with different contours (as a closed circle surrounding the DHT in a time window $\left[T-\Delta T,T+\Delta T\right]$) drawn from the time-averaged flow field $\overline{\varphi}=\frac{1}{T}\int_{0}^{T}\varphi_{t}dt$ (shown in the left panel of Fig.~\ref{fig:Illustration-of-shadowing}).
These contours could be chosen as alternate initial contours to account for errors in the location of the hyperbolic points at the initial time.
As shown in the left panel of Fig.~\ref{fig:Illustration-of-shadowing}, the two large-scale coherent vortices are captured from the initial contours from the averaged field.
We observe that the contours deform from their initial shape and track the shape of the SUMs as long as the time interval $\Delta T$ is taken long enough.
This shows that different choices for the initial contours will lead to qualitatively similar manifolds that demonstrate comparable flow structures.
This independence of the computed SUMs to the choice of initial states and hence shows the robustness of the chaotic tangle picture for understanding the transition to chaos and resulting spatially intermittent patterning of dissipation and transport in in weakly turbulent flows.

\subsection{Self-consistency of PV transport in weakly chaotic flows}

Here, we remark on the connection between the weakly chaotic Lagrangian dynamics described by \eqref{eq:hamiltonian} and the fully self-consistent flow evolution of the BHW equations \eqref{eq:plasma_model} in the Dimits shift regime.
This falls under the issue of \textit{dynamical consistency}, which refers to the fact that the PV is not a passive scalar, and thus the flow \(\mathbf{u}\) should depend on the evolution of \(q\) in a self-consistent way.
Notably, we argue that the that the particular spatially intermittent patterning of turbulent dissipation and mixing induced by the near-integrability and subsequent chaotic tangles is a key ingredient to how the coherent vortex flows can survive to large amplitude.

To illustrate how dynamical consistency can pose a barrier to the maintenance of coherent flows, consider the modified Hasegawa-Mima model \eqref{eq:plasma_onelayer}.
In the mHM model, the streamfunction is determined entirely by the PV, and hence if the streamfunction is time-periodic, the function \(q(x,y,t)\) must also be time-periodic.
However, for time-periodic flows the PV transport occurs in regions of chaotic advection, where integrability has been lost.
This leads to a seeming inconsistency, as the resulting chaotic advection of PV would typically imply chaotic in time behavior for \(q(x,y,t)\), meaning the streamfunction must also have chaotic in time behavior.
Work in \cite{DelCastilloNegrete1993,Del-Castillo-Negrete2000} pointed out that the Rayleigh-Kuo condition predicts that Rossby waves which are marginally stable to flow shear instability are associated with regions where \(\nabla(q+\kappa x) = 0\).
Thus, dynamical consistency can be maintained for the eigenmode-induced flow if the chaotic mixing occurs in regions of constant PV, where the integral of motion provided by the PV is degenerate, and where chaotic advection would not be reflected in the time behavior of \(q(x,y,t)\).

Here, we discuss how near-integrability plays a complementary role in maintaining the dynamical consistency of the observed coherent vortex flows, which arise primarily from a resistive drift-wave instability mechanism rather than a flow shear instability mechanism.
Recall from the discussion in Section~\ref{sec:coherent_vortices} that the coherent vortex dynamics are dominated mostly by advective PV transport, with density transport playing less of a role due to the nearly-adiabatic nature of the vortices.
Since the Lagrangian trajectories cannot cross the lobe boundaries formed by the SUMs, PV transport and dissipation associated with the vortex flows is primarily mediated by the stretching and folding associated with the turnstile dynamics of the lobes.
Noting that the vortex boundaries consist of stable and unstable manifolds, the lobes then primarily accumulate near the vortex boundaries due to the pushing and pulling behavior associated with the SUMs.
As a result, the mixing and dissipation associated with the lobes is primarily confined near the vortex boundaries and does not usually penetrate deeply into either the nearly-laminar vortex cores or zonal flows.

Due to this spatial localization of turbulent behavior, the PV perturbations associated with the vortex cores are able to remain mostly undisturbed.
These PV perturbations then support the strong coherent vortex flow, which provide the separatrices and X-points from which the SUMs and hyperbolic trajectories arise from in the first place.
This then helps to explains the role of the drift-wave eigenmodes in organizing the flow.
Flows which do not resemble either the eigenmodes or other approximate time-periodic solutions to the equations of motion would not induce nearly-integrable Lagrangian flows, and hence induce stronger mixing and dissipation, presumably preventing energy from accumulating in such fluctuations.
Since the eigenmodes are approximate time-periodic solutions to the equations of motion that induce nearly-integrable Lagrangian flows, they can accumulate significant amounts of energy before they begin to induce turbulent mixing or dissipation.
This picture bears a close resemblance to the self-organized nonlinear wave stability property of large-amplitude Rossby waves described in \cite{cao2023rossby}.
This demonstrates the importance of understanding the link between Lagrangian dynamics and spatiotemporal intermittency in describing the dynamics of turbulence in the Dimits shift regime of BHW turbulence.

\subsection{Role of multiscale fluctuations in mixing}

As a more detailed study of the role in large-scale coherent structure and small-scale fluctuations, we illustrate the relation in the detailed full model involving multiscale fluctuations
with the reconstructed field containing only the leading eigenfunctions by comparing the corresponding manifolds and hyperbolic trajectories.
Here, multiscale refers to dynamics active across the range of the DNS scales, i.e. from the large-scale vortices down to the dissipation scales.
Through this comparison, we aim to gain a better understanding about how the large-scale
coherent structures captured by the leading eigenfunctions interact
with the small-scale fluctuations excited by chaotic advection and other nonlinear processes.

\begin{figure*}
\centering
\includegraphics[width=1.\linewidth]{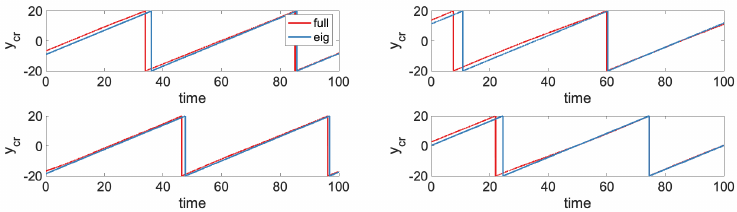}

\caption{The DHTs from the full model and eigenfunctions in the BHW model.\label{fig:DHTs_eig_comp}}
\end{figure*}

\begin{figure*}
\centering
\includegraphics[width=.6\linewidth]{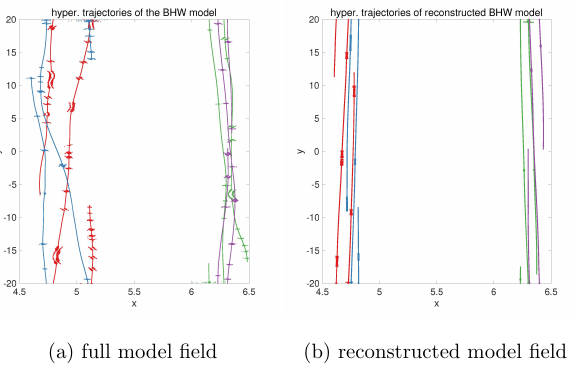}

\caption{Final converged DHTs in the 2-dimensional domain from the full model
flow field and the eigenfunction reconstructed field.\label{fig:DHTs_eig_comp1}}
\end{figure*}

First, the ISPs recovered from the full model and eigenfunction field have been compared in Fig.~\ref{fig:The-flow-potential}.
Focused on the localized region with the corresponding moving speed, the ISPs can be identified at the approximate location using only the eigenfunction field.
Next in Fig.~\ref{fig:DHTs_eig_comp} and \ref{fig:DHTs_eig_comp1}
we plot the DHTs from both the full model simulation data and the
reconstructed eigenfunction field according to the 4 identified ISPs. Clearly, the DHTs from the eigenfunction
fields using only large-scale modes of coherent structures correctly
captures the essential locations of the hyperbolic trajectories and the constant moving speed along the $y$ direction during
their time evolutions. On the other hand, compared with the DHTs recovered
from the full model simulation data, the detailed trajectory paths for local
fluctuations are missed. This is not surprising since the leading
eigenfunctions do not contain such information in small-scale fluctuations.
Still, these becomes essential when we want to analyze the flow development
to full turbulence.

\begin{figure*}
\centering
\includegraphics[width=1.\linewidth]{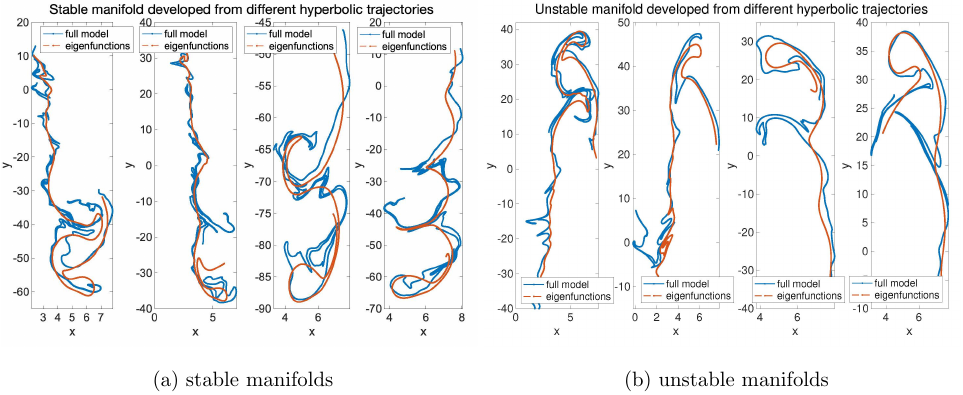}

\caption{Comparison of the stable and unstable manifolds computed from the
full model and reconstructed field from the leading eigenfunctions.\label{fig:mani_comparison}}
\end{figure*}

Finally in Fig.~\ref{fig:mani_comparison}, we compare the stable and unstable
manifold computed from the full model data and the reconstructed field
using eigenfunctions. In this way, we can understand more on the generating
mechanism for the persistent zonal structure and the suppressed zonal
flux transport. The manifolds are transported along the $y$ direction
by the zonal velocity $v$. We do not normalize them back the $\left[-20,20\right]$
domain for a more clear comparison. First, it can be seen that the
manifolds from the eigenfunction field successfully track the large-scale
shapes of the manifolds especially the `lobe' structures. On the other
hand, the small-scale detailed structures are missed in the eigenfunction
reconstructed manifolds. This is because only the leading eigenfunctions
are used in reconstructing the flow fields. 

\section{Summary and Outlook} \label{sec:discussion}

In summary, this work primarily focused on a computational and theoretical analysis of intermittency using techniques from dynamical systems to study the Dimits shift regime of the flux-balanced Hasegawa-Wakatani (BHW) equations.
Using direct numerical simulations, we demonstrated how BHW flows in this regime primarily organize into zonal flows plus large-amplitude drift-wave vortex flows which take on a strong linearly coherent eigenmode character.
Furthermore, we observed a connection between the topology of the flow induced by the eigenmodes and the filamentary structure of the turbulent intermittency observed in the DNS.

Motivated by these observations, we discussed the connection between PV conservation, fluid parcel relabeling symmetries, and integrability in systems which are possibly non-Hamiltonian.
Eigenmodes were shown to formally act as ``nearly-integrable'' perturbations to the zonal flow Lagrangian dynamics.
Then, we discussed the route to chaos through separatrix splitting into transversally intersecting stable and unstable manifolds, forming chaotic tangles.
Using a theory of distinguished hyperbolic trajectories relevant to finite-time aperiodic flows, we computed SUMs in the DNS flow fields and showed how the structure of intermittency in the dissipation and transport of PV originated from the turnstile dynamics of lobes associated with chaotic tangles.
Finally, we argued how the particular spatially intermittent patterning of PV dissipation and transport was a key ingredient in the survival of the coherent vortices to large amplitude.

One physical interpretation of these results is that the Dimits shift regime in BHW appears to be dominated by mesoscale ``convective rolls'' associated with the zonal flows plus coherent vortex flows.
Rather than having their structure determined by interaction with the boundaries of the domain \(\sim L\) or by the microscopic wave physics \(\sim \rho_s\), the structure of the rolls here is determined by an eigenmode envelope equation associated with the `waveguide' structure formed by corrugations in the zonal PV and density profiles.
The spatiotemporally localized transport events associated with the lobe dynamics then bear some kinematic resemblance to the plumes in Rayleigh-B\'{e}nard turbulence, which have also been linked to chaotic tangles formed by the the intersection of SUMs associated with convection rolls \cite{Camassa1991}.

An open issue we do not tackle here is an explicit calculation of how the transport and dissipation mediated by the lobes and chaotic tangles affects the eigenmode and zonal flow amplitudes and phases.
Thus for example, it is not clear if the coherent vortices play a causal role in maintaining the strong Dimits shift in the BHW equation, or if they are simply structures that happens to form in this regime.
This is a potential area for future work, as there are explicit analytical formulas using the framework of Melnikov theory that can help compute the total transport mediated by the lobes \cite{malhotra1998geometric,Balasuriya2016}.
Such work could help establish a self-consistent picture of how the Dimits shift regime evolves in the BHW equations.

Furthermore, while we did not explicitly pursue the study of three-dimensional and/or non-Hamiltonian flows in this work, the discussions in section \ref{sec:near_integrability} are applicable to generic convective nonlinearities, so long as there are enough symmetries or conserved quantities.
In particular, the formal near-integrability property stems from a subset of the equations of motion for the fully self-consistent flow field, suggesting that many types of approximate solutions, not just eigenfunctions, could be relevant to intermittent flows which have the spatiotemporal coexistence of both laminar and turbulent regions.
The notion of stable/unstable manifolds and their connection to chaos have been pursued for three-dimensional time-dependent flows \cite{Fountain1998,Haller2015}.
It would be interesting to see if the chaotic tangle structure persists in simulations of plasma turbulence with more realistic physics and geometries, such as in gyrofluid or even gyrokinetic simulations.

One final remark is that the dynamical systems techniques used here are only sensitive to the evolution of structures due to the convective nonlinearity.
Other effects such as diffusion or resistivity due to particle collisions can only affect these structures through their impact on the velocity fields.
In the BHW system, this is important for characterizing the transport of density, as the adiabaticity \(\alpha\) plays a large role in distinguishing the transport of density from the transport of PV.
Additionally, since the lobe structures are stretched exponentially fast both backwards and forwards in time, the lobes are both initially born and eventually dissipate in scales where diffusive effects dominate.
Thus, a complete description of structures in these flows requires an understanding of how the convective nonlinearity interacts with dissipative and other non-advective effects.

One potential way to incorporate diffusive or resistive effects is through the incorporation of stochastic terms into the characteristic equations.
Solutions to advection-diffusion equations with source/sink terms can be recast into questions about expectations over realizations of solutions to stochastic differential equations via the Feynman-Kac formula.
This has been applied, for example, in the Constantin-Iyer formulation of the Navier-Stokes equations \cite{Constantin2008}, which gives a description to solutions of the Navier-Stokes equations in terms of stochastic Lagrangian trajectories.
This had led to the development of stochastic least-action principles for fluid equations \cite{Eyink2010}, as well as equivalents to the fluid parcel relabeling symmetry and Kelvin's circulation theorem relevant to viscous flows.
Future work can aim to generalize the discussions of near-integrability and chaotic tangles to the stochastic case, paving the way for a unified Lagrangian view of intermittent turbulent structures in the presence of driving and dissipation.

\begin{acknowledgments}
The authors thank Antoine Cerfon and Jalal Shatah for many discussions that inspired this research.
N.M.C. acknowledges support from the Simons Collabration on Wave Turbulence, grant \#617006. 
The research of D.Q. is partially supported by the start-up funds  provided by Purdue University.
\end{acknowledgments}

\section*{Author Declarations}

\subsection*{Conflict of Interest}
The authors have no conflicts to disclose.

\subsection*{Author Contributions}
\textbf{Norman M. Cao:} conceptualization (equal), formal analysis (equal), methodology (equal), software (equal), writing -- original draft (equal), writing -- revision (equal).
\textbf{Di Qi:} conceptualization (equal), formal analysis (equal), methodology (equal), software (equal), writing -- original draft (equal), writing -- revision (equal).

\section*{Data Availability Statement}
The data that support the findings of this study are available from the corresponding author upon reasonable request.

\appendix

\section{Methodology for Proper Orthogonal Decomposition} \label{app:pod}

To compute the POD modes, we take a sequence of snapshots for the vorticity $\omega=\nabla^{2}\varphi$ and density $n$ field in $s=200$ consecutive time steps with $\Delta t=0.5$ (that is, a time window of size $T=100$).
We then rewrite the BHW model solution \eqref{eq:plasma_balance1} and \eqref{eq:plasma_balance2} together as a vector field $\vec{\phi}=\left(q,n\right)$
\begin{equation}
\vec{\phi}\left(\mathbf{x},t_{m}\right)=\sum_{k}a_{k}\left(t_{m}\right)\vec{u}_{k}\left(\mathbf{x}\right),\label{eq:pod}
\end{equation}
with the spatial variable $\mathbf{x}=\left(x,y\right)$.
Above, $a_{k}$ is the time coefficient and $\vec{u}_{k}$ defines the POD basis of spatial modes.
We then introduce the spatial inner product based on the total energy of the system, that is,
\begin{align*}
\left\langle \vec{\phi}_{1},\vec{\phi}_{2}\right\rangle  & =\frac{1}{2}\int\left(-\varphi_{1}\zeta_{2}+n_{1}n_{2}\right)d\mathbf{x}\\
 & =\frac{1}{2}\sum_{\mathbf{k}}k^{2}\hat{\varphi}_{1,\mathbf{k}}\hat{\varphi}_{2,\mathbf{k}}^{*}+\hat{n}_{1,\mathbf{k}}\hat{n}_{2,\mathbf{k}}^{*}.
\end{align*}
Accordingly, we can define the time correlation matrix $C\in\mathbb{R}^{s\times s}$ based on the energy inner product
\[
C_{ij}\equiv C\left(t_{i},t_{j}\right)=
\left\langle\vec{\phi}\left(\mathbf{x},t_{i}\right),\vec{\phi}\left(\mathbf{x},t_{j}\right)\right\rangle .
\]
Notice that \(C\) is a real-valued positive semi-definite symmetric matrix, so its eigenvectors will be orthogonal and the eigenvalues non-negative.
The $k$-th eigenvector of the correlation matrix $C$ gives the orthonormal system of temporal vectors $\mathbf{a}_{k}=\left(a_{k}\left(t_{1}\right),\cdots,a_{k}\left(t_{s}\right)\right)^{T}\in\mathbb{R}^{s}$
which satisfy
\[
C\mathbf{a}_{k}=\lambda_{k}\mathbf{a}_{k},\quad\mathbf{a}_{k}^{T}\mathbf{a}_{l}=\delta_{kl}.
\]

The time-integrated energy of the snapshots \(E\) can be computed in terms of the matrix \(C\) as
\begin{equation*}
    E = \sum_{i=1}^{s} \left\langle\vec{\phi}\left(\mathbf{x},t_{i}\right),\vec{\phi}\left(\mathbf{x},t_{i}\right)\right\rangle
    = \operatorname{Tr} C
    = \sum_{k=1}^{s} \lambda_k.
\end{equation*}
Thus, the \(\lambda_k\) gives the time-integrated energy associated with the temporal vector \(\mathbf{a}_k\).
The \(\lambda_k\) give a ranking to the POD modes based on their energy content observed in the DNS.
Using the orthonormality of the eigenvectors $\mathbf{a}_{k}$, the spatial POD modes are computed by taking the weighted time average
\begin{equation}
\vec{u}_{k}\left(\mathbf{x}\right)=\sum_{m=1}^{s}\vec{\phi}\left(\mathbf{x},t_{m}\right)a_{k}\left(t_{m}\right).\label{eq:pod_mode}
\end{equation}
This also implies the projection property for the POD modes thus infers the orthogonality in the modes $\vec{u}_{k}$
\[
\left\langle \vec{\phi}\left(\mathbf{x},t_{m}\right),\lambda_{k}^{-1}\vec{u}_{k}\left(\mathbf{x}\right)\right\rangle =\lambda_{k}^{-1}\sum_{l}a_{l}\left(t_{m}\right)\left\langle \vec{u}_{l}\left(\mathbf{x}\right),\vec{u}_{k}\left(\mathbf{x}\right)\right\rangle =a_{k}\left(t_{m}\right).
\]

\section{Lagrangian Flow Integrability in Generic Hydrodynamic Systems} \label{app:integrability}

In this appendix, we follow the presentation \cite{Zung2016} to give a precise definition of Generalized Liouville integrability.
As before, we will not distinguish between vector fields and their first-order differential operators, and furthermore we will assume all functions are smooth.
Vector fields are also assumed to be autonomous, which can be done by extending the phase space to include time if necessary.

\begin{defn} \label{def:integrability}
    A vector field \(X_1\) on a manifold \(M\) is said to be \textit{integrable of type} \((p,q)\), where \(p \ge 1\), \(q \ge 0\), and \(p+q = \dim M\), if there exist \(p\) vector fields \(X_1, X_2, ..., X_p\) and \(q\) functions \(F_1, ..., F_q\) on \(M\) which satisfy the following conditions:
    \begin{enumerate}
        \item the vector fields \(X_1, ..., X_p\) commute pairwise, that is \([X_i, X_j] = 0\) for all \(i,j\);
        \item the functions \(F_1,..., F_q\) are common integrals of motion of \(X_1,..., X_p\), that is \(X_i(F_j) = 0\) for all \(i,j\);
        \item \(X_1, ..., X_p\) are linearly independent almost everywhere (i.e. at all points on \(M\) except possibly on a set of measure zero);
        \item \(\nabla F_1, ..., \nabla F_p\) are linearly independent almost everywhere.
    \end{enumerate}
    Under these conditions, we will also say \((X_1, ..., X_p, F_1, ..., F_q)\) is an \textit{integrable system of type} \((p,q)\).
\end{defn}
Often \(X_1\) is taken to be some physical flow, while \(X_2,...,X_p\) are the commuting symmetries.
\begin{defn}
    We say a common level set of the integrals of motion \((F_1,...,F_q) = (z_1,...,z_q)\) is a \textit{regular level set} if \(X_1,...,X_p\) are linearly independent and \(\nabla F_1, ..., \nabla F_q\) are linearly independent everywhere on the level set.
\end{defn}
This definition of regular level sets can be interpreted as a local independence and non-degeneracy condition for the integrals of motion.
Together, these two definitions are sufficient for a rigorous statement of the Generalized Liouville theorem, see Theorem 2.1 in \cite{Zung2016}.
Note that the tuple \((D_t, L_q, q)\) in Proposition \ref{prop:2d_integrability_text} is an integrable system of type \((2,1)\).

Now, we introduce two brackets which allow for the concise statement and proof of two propositions regarding the Liouville integrability of certain hydrodynamic flows.
Given scalar functions \(f,g\) of \(x,y\) and possibly \(t\), we note the Poisson bracket can be written in terms of the Jacobian determinant
\begin{equation*}
    \{f,g\} := \partial_x f \partial_y g - \partial_y f \partial_x g = \det \frac{\partial(f,g)}{\partial(x,y)}.
\end{equation*}
We can extend this to scalar functions \(f,g,h\) of \(x,y,z\) and possibly \(t\) using the Nambu bracket \cite{Nambu1973},
\begin{equation*}
    \{f,g,h\} := \det\frac{\partial(f,g,h)}{\partial(x,y,z)} = (\nabla g \times \nabla h) \cdot \nabla f.
\end{equation*}
Note that the vector field in \eqref{eq:lq} can be written as \(L_q := \{\cdot,q\}/h\).

\begin{prop}[Lagrangian flow integrability in 3d hydrodynamic systems] \label{prop:int3d}
	Let \((\mathbf{u}, \rho, q, \eta)\) be a tuple consisting of a velocity field \(\mathbf{u}(\mathbf{x},t)\), a positive scalar function \(\rho(\mathbf{x},t) > 0\), and two scalar functions \(q(\mathbf{x},t),\eta(\mathbf{x},t)\).
	
	Suppose \((\mathbf{u}, \rho, q, \eta)\) satisfy a mass continuity equation and two scalar conservation equations for \(q\) and \(\eta\),
	    
	\begin{subequations} \label{eq:3d_integrability}
        \begin{gather} 
            D_t \rho + \rho (\nabla \cdot \mathbf{u}) = 0, \\
            D_t q = 0, \\
            D_t \eta = 0.
        \end{gather}
    \end{subequations}
	
	Then, the vector field \(L_{q,\eta}\) defined by
	\begin{equation}
		L_{q,\eta} := \left(\rho^{-1} \nabla q \times \nabla \eta\right) \cdot \nabla = \frac{\{\cdot,q,\eta\}}{\rho}
	\end{equation}
	satisfies \([D_t,L_{q,\eta}] = 0\). In particular, if \(\nabla q\) and \(\nabla \eta\) are linearly independent except possibly on a set of measure zero, then \((D_t, L_{q,\eta}, q, \eta)\) is an integrable system of type \((2,2)\).
\end{prop}

The vector fields \(L_q\) and \(L_{q,\eta}\) are both interpreted as generators for the fluid parcel relabeling symmetry.
In the 3d case, \(\eta\) would usually be a specific entropy or potential density/temperature type of scalar.
The form of \(L_{q,\eta}\) suggests that only fluid parcels on surfaces of constant \(\eta\) can be exchanged with each other under the fluid parcel relabeling symmetry.
Furthermore, it can be shown that these flows preserve the densities \(h\) and \(\rho\) (in the sense of acting as a Lie derivative on an area/volume form) in their respective cases.

The proofs of Proposition \ref{prop:2d_integrability_text} and \ref{prop:int3d} are extremely similar.
The key step in either proof relies on the following identities, which can be verified using the product rule or using a computer algebra system:
\begin{subequations}
	\begin{gather}
		D_t \{f,g\} = \{ D_t f, g\} + \{f, D_t g\} - (\nabla \cdot \mathbf{u}) \{f,g\}, \\
		D_t \{f,g,h\} = \{D_t f, g, h\} + \{f, D_t g, h\} + \{f, g, D_t h\}  - (\nabla \cdot \mathbf{u}) \{f,g,h\}.
	\end{gather}
\end{subequations}
These identities arise from the fact that \(\{f,g,h\}\) is the oriented volume enclosed by the parallelepiped spanned by \(\nabla f, \nabla g, \nabla h\).
The differential change in volume following the flow \(D_t\) will be given by the sum of the differential changes of volume due to derivatives of \(f, g, h\) following the flow, compensated by the change in volume due to the divergence of \(\mathbf{u}\).
A similar picture holds for \(\{f,g\}\), which is the oriented area enclosed by the parallelogram spanned by \(\nabla f, \nabla g\).
Note the proof of these propositions is very similar in spirit to the proofs in \cite{Mezic1994,Haller1998} and can also be seen as an application of the method of inverse Jacobi multipliers \cite{Berrone2003}.

\begin{proof}
	(Proposition \ref{prop:2d_integrability_text}). Suppose \((\mathbf{u}, h, q)\) satisfy the hypotheses of Proposition \ref{prop:2d_integrability_text}. Then, for any scalar function \(f\), we have
	\begin{align*}
		D_t (L_q (f)) &= D_t \left(\frac{\{f,q\}}{h}\right) \\
		&= -\frac{D_th + h (\nabla \cdot \mathbf{u})}{h} \frac{\{f,q\}}{h} + \frac{\{D_tf,q\}}{h}  + \frac{\{f,D_t q\}}{h} \\
		&= \frac{\{D_tf,q\}}{h} \\
		&= L_q(D_t(f))
	\end{align*}
	which shows that \([D_t,L_q] = 0\).
	
	\(L_q (q) = \frac{\{q,q\}}{h} = 0\), so \(q\) is a common integral of motion for \(D_t, L_q\). Then, if \(\nabla q \neq 0\) except possibly on a set of measure zero, \(L_q\) and \(D_t\) are linearly independent (i.e. since \(D_t\) has a \(\partial_t\) term and \(L_q\) does not). Thus, \((D_t, L_q, q)\) is an integrable system of type \((2,1)\).
	
	(Proposition \ref{prop:int3d}). Suppose \((\mathbf{u}, \rho, q, \eta)\) satisfy the hypotheses of Proposition \ref{prop:int3d}. Then, for any scalar function \(f\) we have
	\begin{align*}
		D_t(L_{q,\eta}(f)) &= D_t \left(\frac{\{f,q,\eta\}}{\rho}\right) \\
		&= -\frac{D_t \rho + \rho (\nabla \cdot \mathbf{u})}{\rho} \frac{\{f,q,\eta\}}{\rho} + \frac{\{D_tf,q,\eta\}}{\rho} + \frac{\{f,D_tq,\eta\}}{\rho} + \frac{\{f,q,D_t\eta\}}{\rho} \\
		&= \frac{\{D_tf,q,\eta\}}{\rho} \\
		&= L_{q,\eta}(D_t(f))
	\end{align*}
	which shows that \([D_t,L_{q,\eta}] = 0\).
	
	\(\{q,q,\eta\} = \{\eta,q,\eta\} = 0\), so \(L_{q,\eta}(q) = L_{q,\eta}(\eta) = 0\). Thus, \(q,\eta\) are common integrals of motion for \(D_t, L_{q,\eta}\). Then, if \(\nabla q\) and \(\nabla \eta\) are linearly independent, \(\nabla q \times \nabla \eta\) is non-zero. Thus, \(L_{q,\eta}\) and \(D_t\) are linearly independent, so \((D_t, L_{q,\eta}, q, \eta)\) is an integrable system of type \((2,2)\).
\end{proof}

To show how to use these propositions in practice and show how subtleties can arise, we provide an explicit proof of Theorem \ref{thm:chm_integrability}.

\begin{proof}
    (Theorem \ref{thm:chm_integrability}) Let \((q(x,y,t),n(x,y,t))\) be a smooth time-periodic solution to the BHW equations \eqref{eq:plasma_model} with \(\mu=0\) on a doubly-periodic domain. Let \(\mathbf{u}\) be the corresponding \(E \times B\) flow velocity.
    
    We extend the domain in \(x\) to the entire real line so that \(\kappa x\) is a singly-valued function, and work in the extended phase space isomorphic to \(M := \mathbb{R} \times \mathbb{T} \times \mathbb{T}\), where \(\mathbb{T}\) is a 1-torus. Since \(q\) is time-periodic, it will be a singly-valued function of time. Then by the BHW equations and the assumption that \(\nabla(q+\kappa x) \neq 0\) except possibly on a set of measure zero, the tuple \((\mathbf{u}, 1, q+\kappa x)\) satisfies the requirements of proposition \ref{prop:2d_integrability_text} on the manifold \(M\).

    Since \(q\) is smooth and triply-periodic in \((x,y,t)\), it will be bounded by some finite \(K := \sup_{x,y,t}|q(x,y,t)|\). Then, each \(c\) level set of \(q+\kappa x\) will be contained within the compact set \([(c-K)/\kappa, (c+K)/\kappa] \times \mathbb{T} \times \mathbb{T} \subset M\). Since level sets of continuous functions are closed, every level set of \(q+\kappa x\) is a closed subset of a compact set, and hence is also compact. Thus, all of the requirements of the Generalized Liouville theorem are satisfied for the Lagrangian flow associated to \(\mathbf{u}\).
\end{proof}

Now, as an example application of Proposition \ref{prop:int3d}, consider 3d adiabatic ideal MHD on a domain without boundary, i.e. \(\mathbb{R}^3\) or a triply periodic domain.
The physical variables are the fluid velocity \(\mathbf{u}\), mass density \(\rho\), specific entropy \(\eta\), and the magnetic field \(\mathbf{B}\).
These satisfy
\begin{subequations}\label{eq:mhd}
    \begin{gather}
        D_t \mathbf{u} = -\frac{1}{\rho}\nabla p(\rho,\eta) + \frac{1}{\rho} \mathbf{J} \times \mathbf{B}, \\
        D_t \rho = - \rho \nabla \cdot \mathbf{u}, \label{eq:mhd_mass} \\
        D_t \eta = 0, \\
        \partial_t \mathbf{B} = \nabla \times (\mathbf{u} \times \mathbf{B}). \label{eq:mhd_induction}
    \end{gather}
\end{subequations}
The divergence-free condition \(\nabla \cdot \mathbf{B} = 0\) is preserved by the dynamics.

It is well known that the induction equation \eqref{eq:mhd_induction} can be combined with the density equation \eqref{eq:mhd_mass} to get
\begin{equation*}
    D_t (\mathbf{B}/\rho) = (\mathbf{B}/\rho) \cdot \nabla \mathbf{u}.
\end{equation*}
This equation is equivalent to the commutator relation \([D_t,L_{\mathbf{B}/\rho}] = 0\) for the vector field \(L_{\mathbf{B}/\rho} := (\mathbf{B}/\rho) \cdot \nabla\).
This vector field can be used to define an electron potential canonical vorticity,
\begin{equation} \label{eq:mhd_pv}
    q := \frac{\mathbf{B} \cdot \nabla \eta}{\rho} = L_{\mathbf{B}/\rho}[\eta].
\end{equation}
This is a well-known invariant of MHD satisfying \(D_t q = 0\), see e.g. \cite{Padhye1996} for a discussion of how it relates to fluid parcel relabeling symmetries.
Now, we state the following theorem, which is a direct consequence of Proposition \ref{prop:int3d}:
\begin{thm}[Lagrangian flow integrability of time-periodic solutions to MHD]
    Suppose \((\mathbf{u},\rho,\eta,\mathbf{B})\) is a time-periodic solution to the 3d ideal adiabatic MHD equations \eqref{eq:mhd} with \(\mathbf{B} \neq 0\) almost everywhere. Then, the tuple \((\mathbf{u},\rho,q,\eta)\) with \(q\) defined by equation \eqref{eq:mhd_pv} satisfies the requirements of Proposition \ref{prop:int3d}. In particular if \(q\) and \(\eta\) have compact level sets, and \(\nabla q\) and \(\nabla \eta\) are linearly independent (hence non-zero) almost everywhere, then the Lagrangian flow \(D_t\) will be Liouville integrable in the generalized sense.
\end{thm}

This gives an example of an integrability theorem for the Lagrangian flow in a time-periodic three-dimensional compressible system.
Note that it may frequently occur in practice that the invariants here are degenerate, i.e. \(\nabla \eta = 0\) or \(q=0\).
In this case there may be other invariants or commuting vector fields that can be used to derive different integrability conditions.
However, we do not pursue this here.

We finish this section by remarking on the assumption of smoothness in all the fields.
While this requirement can likely be relaxed, non-smooth behavior in the velocity fields can lead to unexpected behavior.
For example, velocity fields with a sufficiently shallow power-law decay in their energy spectrum (such as fields that are H\"{o}lder but not Lipschitz continuous) have recently been shown to be linked with anomalous dissipation in the inviscid limit via the Onsager conjecture \cite{Eyink2008,PhilipIsett2018,DeLellis2019}.
Such roughness in the velocity field also can also lead to a breakdown of the uniquness of solutions to the ODE \eqref{eq:hamiltonian}.
This loss of uniquess has been called ``spontaneous stochasticity'' in the physics literature, which has also been linked to anomalous dissipation \cite{Chaves2003,Drivas2017}.
Thus, even in the absence of viscosity, it may be possible for the fluid parcel relabeling symmetries generated by \(L_q\) and \(L_{q,\eta}\) to experience ``anomalous symmetry breaking'' for sufficiently rough fields, in which case they may not guarantee Liouville integrability.

\bibliography{refs}

\end{document}